\documentclass[prx,twocolumn,superscriptaddress,showpacs,amsmath,amstex,amssymb,citeautoscript]{revtex4-1}
\pdfoutput=1 

\usepackage{dcolumn}
\usepackage{bm}
\usepackage[T1]{fontenc}
\usepackage{braket}
\usepackage{graphicx}
\usepackage{xcolor}
\usepackage{lipsum}
\usepackage[normalem]{ulem}
\usepackage{amsmath}
\usepackage{amsfonts}
\usepackage{amssymb} 
\usepackage{color}
\usepackage[percent]{overpic}
\usepackage{soul} 
\usepackage{amssymb}
\usepackage{wasysym}
\usepackage{dsfont}
\usepackage{float}
\usepackage[mathscr]{euscript}
\usepackage{ifthen}
\usepackage{csquotes}
\usepackage{appendix}  
\usepackage{verbatim}

\begin{document}

\title{Transverse spin dynamics in the anisotropic Heisenberg model \\ realized with ultracold atoms}

\author{Paul Niklas Jepsen}
\affiliation{Department of Physics, Massachusetts Institute of Technology, Cambridge, MA 02139, USA}
\affiliation{Research Laboratory of Electronics, Massachusetts Institute of Technology, Cambridge, MA 02139, USA}
\affiliation{MIT-Harvard Center for Ultracold Atoms, Cambridge, MA, USA}

\author{Wen Wei Ho}
\affiliation{MIT-Harvard Center for Ultracold Atoms, Cambridge, MA, USA}
\affiliation{Department of Physics, Harvard University, Cambridge, MA 02138, USA}
\affiliation{Department of Physics, Stanford University, Stanford, CA 94305, USA}

\author{Jesse Amato-Grill}
\affiliation{Department of Physics, Massachusetts Institute of Technology, Cambridge, MA 02139, USA}
\affiliation{Research Laboratory of Electronics, Massachusetts Institute of Technology, Cambridge, MA 02139, USA}
\affiliation{MIT-Harvard Center for Ultracold Atoms, Cambridge, MA, USA}
\affiliation{QuEra Computing Inc., Boston, MA 02135, USA}

\author{Ivana Dimitrova}
\affiliation{Department of Physics, Massachusetts Institute of Technology, Cambridge, MA 02139, USA}
\affiliation{Research Laboratory of Electronics, Massachusetts Institute of Technology, Cambridge, MA 02139, USA}
\affiliation{MIT-Harvard Center for Ultracold Atoms, Cambridge, MA, USA}
\affiliation{Department of Physics, Harvard University, Cambridge, MA 02138, USA}

\author{Eugene Demler}
\affiliation{MIT-Harvard Center for Ultracold Atoms, Cambridge, MA, USA}
\affiliation{Department of Physics, Harvard University, Cambridge, MA 02138, USA}

\author{Wolfgang Ketterle}
\affiliation{Department of Physics, Massachusetts Institute of Technology, Cambridge, MA 02139, USA}
\affiliation{Research Laboratory of Electronics, Massachusetts Institute of Technology, Cambridge, MA 02139, USA}
\affiliation{MIT-Harvard Center for Ultracold Atoms, Cambridge, MA, USA}
 
\begin{abstract}
In Heisenberg models with exchange anisotropy, transverse spin components are not conserved and can decay not only by transport, but also by dephasing. Here we utilize ultracold atoms to simulate the dynamics of 1D Heisenberg spin chains, and observe fast, local spin decay controlled by the anisotropy. Additionally, we directly observe an effective magnetic field created by superexchange which causes an inhomogeneous decay mechanism due to  variations of lattice depth between chains, as well as dephasing within each chain due to the twofold reduction of the effective magnetic field at the edges of the chains and due to fluctuations of the effective magnetic field  in the presence of mobile holes. The latter is a new coupling mechanism between holes and magnons.  All these dephasing mechanisms, corroborated by extensive numerical simulations, have not been observed before with ultracold atoms and illustrate basic properties of the underlying Hubbard model. 
\end{abstract}

\maketitle

\section{Introduction}

The famous Heisenberg Hamiltonian, also called the Heisenberg-Dirac-van Vleck Hamiltonian \cite{Heisenberg1926,Dirac1926,VanVleck1932}, describes localized particles on a lattice  interacting via spin  exchange couplings. Despite its apparent simplicity, it serves as a paradigmatic model for a host of emergent phenomena, such as ferromagnetism (due to Coulomb exchange, also called potential or direct exchange), antiferromagnetism (due to kinetic exchange from tunneling, also called superexchange) \cite{assa}, spin-glass physics \cite{RevModPhys.58.801}, as well as  exotic states of matter like topologically ordered quantum spin liquids \cite{Savary_2016}. The dynamics of such models is also very rich and multi-faceted, and is under active, intense investigation. For example, in one dimension, Heisenberg spin models (with spin quantum number $S\,{=}\,1/2$) have the special property of being integrable, whereby stable quasiparticles exist at all temperatures. This gives rise to a breakdown of simple hydrodynamics with accompanying varied spin transport behaviors \cite{Vasseur2016, Bertini2020, Ljubotina2017, Gopalakrishnan2019, Ilievski2018}. Understanding this has led to the recent development of a theory of generalized hydrodynamics  \cite{Bertini2016_GHD,CastroAlvaredo2016_GHD}. In higher-dimensions, the interplay of spontaneous symmetry-breaking can lead to long-lived, metastable, prethermal states  in addition to the onset of regular spin diffusion \cite{Knap2015, PhysRevLett.125.230601, rodrigueznieva2020transverse} or even turbulent relaxation with universal scaling of spin-spin correlations \cite{rodrigueznieva2020turbulent}.

Ultracold atoms in  optical lattices  form  an ideal platform to realize Heisenberg spin models and probe their   dynamics in a controlled fashion \cite{Bloch2017_OpticalLatticesReview}. In deep lattices  where atoms are localized and Mott insulators form \cite{Jaksch1998_BosonsInOpticalLattices}, superexchange processes   via second-order tunneling yield  effective Heisenberg spin models, with potential tunability of the strength, sign, and anisotropy of the spin-exchange interactions \cite{Svistunov2003_CounterflowSF, Duan2003_ControllingSpinExchange, GarciaRipoll2003_BosonsInOpticalLattice, Altman2003_TwoComponentBosons}. Until very recently, all experimental studies addressed the special case of an isotropic Heisenberg model \cite{Bloch2013_SingleSpin, Bloch2013_BoundMagnons, Bloch2014_SpinHelix,Zwierlein2019_SpinTransport, Zwierlein2016_FermiMottInsulator, Greiner2016_FermiSpinCorrelations, Greiner2017_FermiAntiferromagnet}. However, in \cite{Jepsen2020_SpinTransport}, we showed how to overcome this limitation and implemented Heisenberg models with {\it tunable} anisotropy of the nearest-neighbour spin–spin couplings, by using $^7$Li and varying the interactions through Feshbach resonances.  We were able to show that the anisotropy profoundly changes the nature of transport of longitudinal spin components after a quantum quench from so-called longitudinal spin-helices (see Fig.~\ref{fig:setup}b), and observed ballistic, subdiffusive, diffusive and superdiffusive behavior in different parameter regimes. These results bear some similarities with those of  spin transport close to equilibrium, but strikingly differ in other aspects, prompting the need for further theoretical investigation.

In this paper, we study the relaxation of transverse spin components after quantum quenches from  transverse spin-helices (see Fig.~\ref{fig:setup}a), and observe even more dramatic effects of the anisotropy. In the classical limit, {\it any} transverse spin helix for {\it any} anisotropy is stationary, since the torques exerted by neighboring spins on a given spin exactly cancel (see Appendix \ref{Appendix:Semiclassical}). Therefore, what we study here are the effects of quantum fluctuations on their stability. In contrast to longitudinal spin patterns, which can decay only by transport, transverse spin components can decay also by dephasing. We focus here on two paradigmatic models, which represent complementary spin physics: the XX model, which has only transverse spin-spin couplings and can be mapped to a non-interacting system of fermions, and the XXX model, which has isotropic spin couplings. For the XX model we observe and explain that the decay is faster for spin-helix patterns with longer wavelengths, in contrast to a decay via spin transport, which would entail slower dynamics for longer modulations. We also identify several dephasing mechanisms not discussed before.  For the XXX model, we identify a symmetry-breaking term in the Bose-Hubbard model: an effective magnetic field caused by different scattering lengths for the  spin $\ket{\uparrow}$ and spin $\ket{\downarrow}$ states. This superexchange-induced effective magnetic field is often ignored, as a field that is spatially uniform can be eliminated in the bulk by going into an appropriate co-rotating frame. Here we show that the presence of the effective field is actually significant and gives rise to three additional dephasing mechanisms, resulting in drastically different decay behavior of spin helix patterns with different  orientations.  One is an inhomogeneous effect where the effective magnetic field is non-uniform across different chains in our sample. This can be eliminated with a spin-echo technique. The second  is due to dephasing occurring at the ends of finite chains.  The third   is due to the  presence of mobile holes resulting in a fluctuating effective magnetic field in the bulk, i.e.~a hole-magnon coupling. Our work shows the limitations of a pure spin model in capturing spin dynamics realized with ultracold atoms, and demonstrates the need for a bosonic $\tilde{t}$-$J$ model with hole-magnon couplings in order to reach a more complete description of experiments. The new insight into hole-magnon coupling should be important for  other systems and materials  where such couplings are present, such as high-temperature superconductivity \cite{spalek2007tj,PhysRevB.37.3759,Anderson_2004}.

\section{Experimental methods}

The spin models are implemented with a system of two-component bosons in an optical lattice, which is well-described by the Bose-Hubbard model. These two states (lowest and second-lowest hyperfine states of $^7$Li), labelled $\ket{\downarrow}$ and $\ket{\uparrow}$, form a spin-$1/2$ system.  In the idealized scenario of a Mott insulating regime  at unity filling, bosons cannot tunnel and the effective Hamiltonian for the remaining spin degree of freedoms is given by the spin-1/2 Heisenberg XXZ model \cite{Svistunov2003_CounterflowSF, Duan2003_ControllingSpinExchange, GarciaRipoll2003_BosonsInOpticalLattice, Altman2003_TwoComponentBosons} 
\begin{equation}
	H = \sum_{\langle i j \rangle} \left[ J_{xy} (S_i^x S_j^x + S_i^y S_j^y) + J_z S_i^z S_j^z - \frac{h_z}{2} \left( S^z_i + S^z_j \right) \right], 
    	\label{Heisenberg_eq}
\end{equation}
where $S^\alpha$ ($\alpha\,{=}\,x,y,z$) are spin-1/2 Pauli operators and the sum is over nearest-neighbor pairs of sites $\langle ij\rangle$.  In leading order, one obtains for the transverse coupling $J_{xy}\,{=}\,{-}\,4\tilde{t}^2/U_{\uparrow\downarrow}$, and for the longitudinal coupling   $J_z\,{=}\,4\tilde{t}^2/U_{\uparrow\downarrow}\,{-}\,(4\tilde{t}^2/U_{\uparrow\uparrow}\,{+}\,4\tilde{t}^2/U_{\downarrow\downarrow})$, both mediated by superexchange processes. Here $\tilde{t}$ is the tunneling matrix element between neighbouring sites, while $U_{\uparrow\uparrow}$, $U_{\uparrow\downarrow}$, $U_{\downarrow\downarrow}$ are the on-site interaction energies.  The effective magnetic field strength is $h_z\,{=}\ 4\tilde{t}^2/U_{\uparrow\uparrow}\,{-}\,4\tilde{t}^2/U_{\downarrow\downarrow}$.   Note that the total magnetization $\sum_i S^z_i$ is conserved by the Hamiltonian.

\begin{figure}[t]
	\includegraphics[width=\linewidth,keepaspectratio]{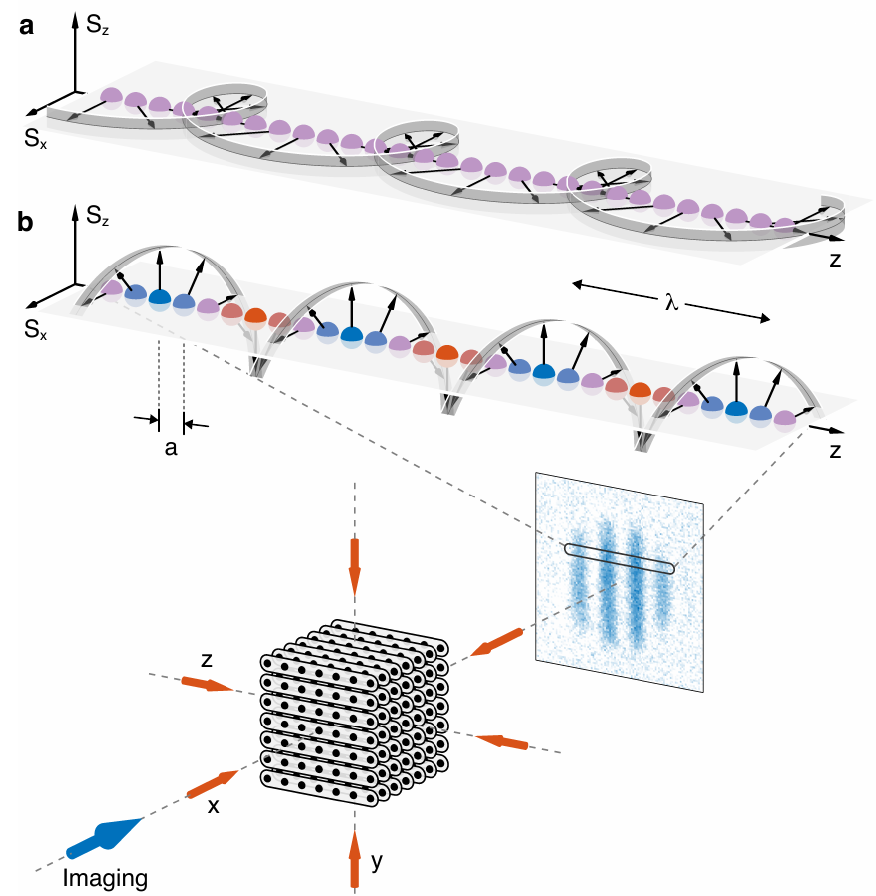}
    	\caption{\textbf{Geometry of the experiment}. \textbf{a} (and \textbf{b}) show the transverse (longitudinal) spin helix  where the spin winds within the $S^x$-$S^y$-plane ($S^z$-$S^x$-plane).  The transverse helix is a pure phase modulation of spin up and down states, whereas the longitudinal helix also involves population modulation. Deep optical lattices along the $x$- and $y$-direction create an array of independent spin chains. The $z$-lattice is shallower and controls spin dynamics along each chain.}
	\label{fig:setup}
	\vspace{-0pt}
\end{figure}

The magnitude of superexchange can be varied over two orders of magnitude by changing the lattice depth, which scales the entire Hamiltonian.  The anisotropy $\Delta\,{:=}\,J_z/J_{xy}$ is controlled via an applied magnetic field which tunes the interactions through Feshbach resonances in the lowest two hyperfine states. In the regime studied here, the transverse coupling is positive ($J_{xy}\,{>}\,0$). The ability to tune the anisotropy over a wide range of values, both positive and negative, allows us to explore dynamics beyond previous experiments \cite{Zwierlein2019_SpinTransport, Brown2015_Superexchange2D, Bloch2013_SingleSpin, Bloch2013_BoundMagnons, Bloch2014_SpinHelix} in which $\Delta\,{\approx}\,1$. 

\begin{figure*}[t]
	\includegraphics[width=\linewidth,keepaspectratio]{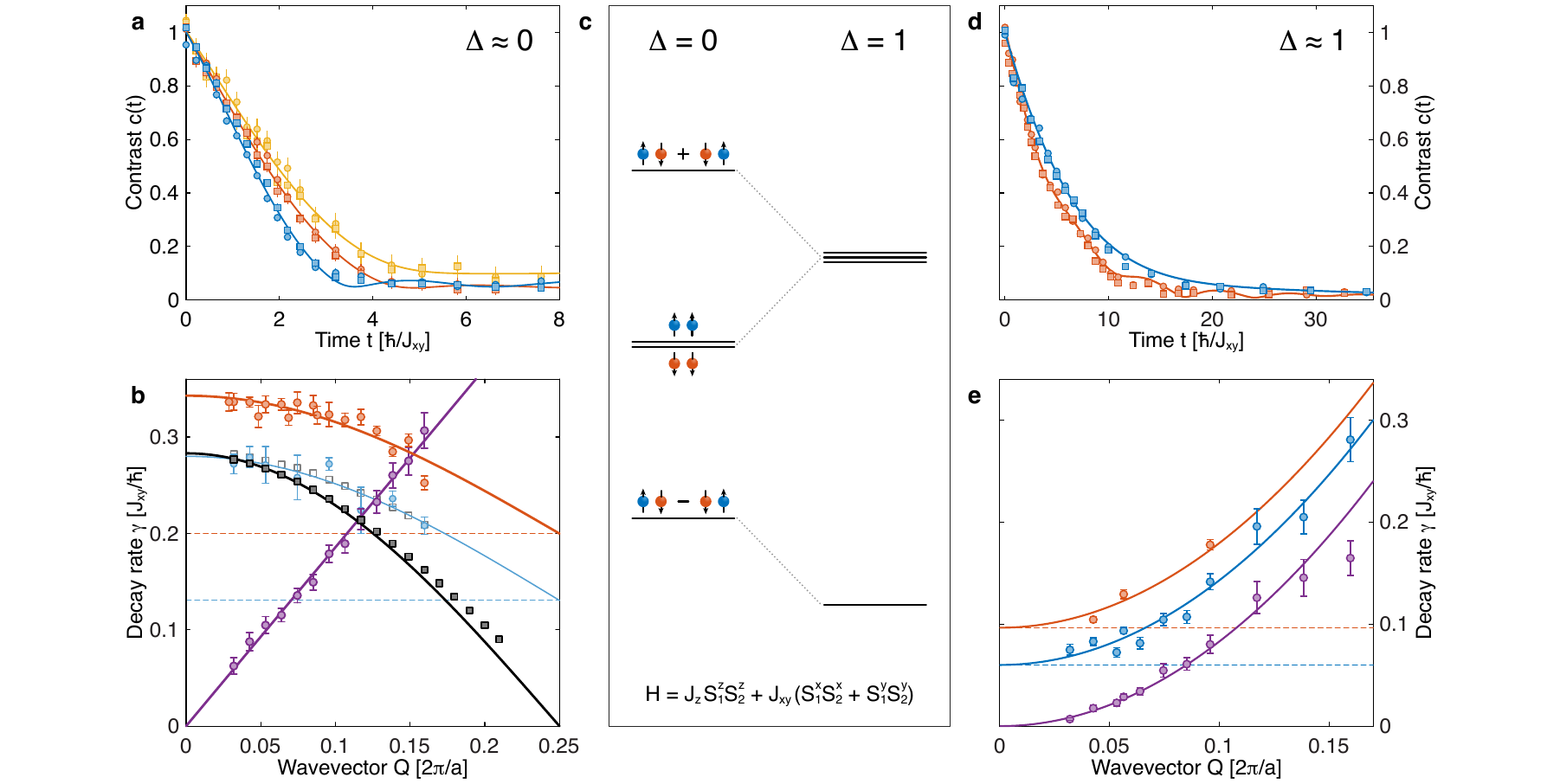}
    	\caption{\textbf{Spin dephasing and spin transport} for the XX model ($\Delta\,{\approx}\,0$, \textbf{a},\textbf{b}) and the isotropic XXX model ($\Delta\,{\approx}\,1$, \textbf{d},\textbf{e}). \textbf{a}, Transverse spin-helix contrast $c(t)$ at $\Delta\,{\approx}\,0$ for $\lambda\,{=}\,31.3\,a$ (blue), $7.2\,a$ (orange), $6.3\,a$ (yellow). The curves taken for different lattice depths $11\,E_R$ ($\circ$) and $13\,E_R$ ($\Box$) collapse when times are rescaled in units of the corresponding spin-exchange times $\hbar/J_{xy}\,{=}\,1.71\,\text{ms}$ ($\circ$) and $4.30\,\text{ms}$ ($\Box$). The transverse spin decays on a time scale of a few spin-exchange times which increases for smaller values of $\lambda\,{=}\,2\pi/Q$. \textbf{b}, The initial decay rate (orange) follows a cosine dependence $\gamma(Q)\,{=}\,\gamma_1 \cos(Qa)\,{+}\,\gamma_0$ (solid orange line) with a constant background rate $\gamma_0\,{=}\,0.20(2)\,J_{xy}/\hbar$ (dashed orange line). This is in strong contrast to the longitudinal spin helix (purple) which shows linear scaling with $Q$ (characteristic of ballistic transport). A spin-echo ($\pi$-pulse at time $t/2$) reduces the background rate to $\gamma_0\,{=}\,0.13(3)\,J_{xy}/\hbar$ (dashed blue line). The black solid squares are numerical results for a single chain and the pure spin model (with $h_z\,{=}\,0$) with a  function $\gamma(Q)\,{=}\,\gamma_1 \cos(Qa)$ (solid line) fitted to the points for  $Q\,{\leq}\,2\pi\,{\times}\,0.075/a$. The grey open squares are numerical results for the $\tilde{t}$-$J$ model with $5\,\%$ hole fraction (see Appendix \ref{subsection:XX_appendix}). \textbf{c}, Eigenenergies of the Heisenberg Hamiltonian for two spins in a double-well potential. For $\Delta\,{\neq}\,1$, the triplet states are split. \textbf{d},  Transverse spin-helix contrast $c(t)$ at $\Delta\,{\approx}\,1$ for $\lambda\,{=}\,10.4\,a$ with (blue) and without (orange) spin-echo ($\pi$-pulse at time $t/2$).  The difference shows the presence of inhomogeneous dephasing. The curves taken for different lattice depths $11\,E_R$ ($\circ$) and $13\,E_R$ ($\Box$) collapse when times are rescaled in units of the corresponding spin-exchange times $\hbar/J_{xy}\,{=}\,2.55\,\text{ms}$ ($\circ$) and $6.42\,\text{ms}$ ($\Box$). \textbf{e}, Decay rates for the transverse (orange, blue) and longitudinal (purple) helix for $\Delta\,{=}\,1$. The blue (with echo) and orange (without echo) curves are fits $\gamma(Q)\,{=}\,DQ^2\,{+}\,\gamma_0$ assuming two contributions to the decay rate: one quadratic term, where $D$ is the diffusion constant (taken from the longitudinal spin decay shown in purple), the other $Q$-independent $\gamma_0$ (shown by the dashed lines). The spin-echo reduces the background decay rate $\gamma_0$ by an amount of $0.036\,J_{xy}/\hbar$ (\textbf{e}) and $0.069\,J_{xy}/\hbar$ (\textbf{b}). These values are consistent with an inhomogeneous dephasing rate which scales linearly with the effect magnetic field $h_z$, which takes the values $h_z\,{=}\,0.89\,J_{xy}$ (\textbf{e}) and $1.43\,J_{xy}$ (\textbf{a}). }
	\label{fig:decayCurves}
	\vspace{-0pt}
\end{figure*}

One-dimensional (1D) chains are created by two perpendicular optical lattices whose depths $V_x$,$V_y\,{=}\,35\,E_R$ are sufficient to prevent superexchange coupling on experimental timescales. A third orthogonal lattice along the $z$-direction with adjustable depth $V_z$  controls the superexchange rate in the chains (Fig.~\ref{fig:setup}). Here $E_R\,{=}\,{h^2/(8ma^2)}$ denotes the recoil energy, where $a\,{=}\,0.532\,\mu\text{m}$ is the lattice spacing, $m$ the atomic mass and $h$ Planck's constant. After preparing a transverse (Fig.~\ref{fig:setup}a) (this work) or longitudinal (Fig.~\ref{fig:setup}b) (studied in our previous work \cite{Jepsen2020_SpinTransport}) spin helix with wavelength $\lambda$ (equivalently, wavevector $Q\,{=}\,2\pi/\lambda$) in each chain \cite{Bloch2014_SpinHelix, Koeh2013_SpinTransportFermiGas, Thywissen2015_LG_effect}, time evolution is initiated by rapidly lowering $V_z$, i.e.~a quench. Dynamics is  then governed by the 1D XXZ model Eq.~\eqref{Heisenberg_eq}.  After an evolution time $t$, the dynamics is frozen by rapidly increasing $V_z$ and the atoms are imaged in the $\ket{\uparrow}$ state via state-selective polarization-rotation imaging with an optical resolution of about 6 lattice sites. For imaging the transverse spin, we apply a $\pi/2$-pulse first, so that we observe the magnetization in the $x$-direction. To distinguish homogeneous from inhomogeneous dephasing, we can use a spin-echo by applying a $\pi$-pulse (with a typical duration of $t_\pi\,{=}\,150\,\mu\text{s}\,{\ll}\,t$) after half of the evolution time $t$.

Integrating the images along the direction perpendicular to the chains yields a 1D spatial profile of the population in the $\ket{\uparrow}$ state, averaged over all spin chains. As in Fig.~\ref{fig:setup}, the spin helix exhibits a sinusoidal spatial modulation of the density of $\ket{\uparrow}$ atoms, observed as a characteristic stripe pattern with a normalized contrast $c(t)$ (see Appendix \ref{Appendix:Analysis} for data fitting methods). During the evolution time $t$ the $100\,\%$ contrast of the initial spin helix decays, and we determine the dependence of $c(t)$ on lattice depth $V_z$, wavevector $Q$, and anisotropy $\Delta$.

In general, we measure the spin dynamics at two or more different lattice depths $V_z$ and verify that the decay curves $c(t)$ collapse when time is rescaled by the corresponding spin-exchange time $\hbar/J_{xy}$, confirming that the dynamics is driven by superexchange processes. Time units are obtained from the experimentally determined lattice depth using an extended Hubbard model (detailed in  \cite{Jepsen2020_SpinTransport}).

\section{Results}

\subsection{XX model}
\label{sec:XXmodel}

We first consider a very anisotropic system by realizing the Heisenberg model tuned to the non-interacting point ($\Delta\,{=}\,0$), and study the decay of transverse spin helices with different wavevectors (Fig.~\ref{fig:decayCurves}a-b).  We find the decays are quick, all having timescales on the order of a few superxchange times $\hbar/J_{xy}$, much faster than the decay for longitudinal spin helices which is driven by  ballistic transport. Importantly, the transverse decay time even {\it decreases for longer wavelengths} of the helix, showing that the decay is not caused by transport, but by dephasing.

Some insight into the fast timescales of transverse decay is obtained  by taking the $Q\,{\to}\,0$ limit, where the initial state becomes a uniform product state. This state is obviously not an eigenstate of the quantum XX model, and is therefore unstable  even in the $Q\,{\to}\, 0$ limit. Further simplification to a two-site (double-well) system allows us to analytically diagonalize the Heisenberg Hamiltonian, which gives a level structure as shown in Fig.~\ref{fig:decayCurves}c (for $h_z\,{=}\,0$). When $\Delta\,{=}\,1$, the transverse spin state $\ket{\rightarrow\rightarrow}\,{:=}\,(\ket{\uparrow}\,{+}\,\ket{\downarrow})(\ket{\uparrow}\,{+}\,\ket{\downarrow})/2$ is an eigenstate of the Hamiltonian as all triplet states are degenerate, and hence does not evolve.  However for $\Delta\,{\neq}\,1$ the degeneracy is lifted and the state $\ket{\rightarrow\rightarrow}$ shows a beat note at the frequency of the energy splitting $J_{xy}\,{\cdot}\,(1\,{-}\,\Delta)/2$. For $\Delta\,{=}\,0$ this indicates a dephasing time for transverse spins on the order of a few superexchange times $\hbar/J_{xy}$, in qualitative agreement with our observations. For many sites, there will be a spectrum of beat frequencies  leading to irreversible dephasing locally.

We can explain the unusual $Q$-dependence of the transverse decay with a semiclassical analysis of   spin dynamics  (see Appendices \ref{Appendix:Semiclassical}, \ref{appendix:dispersion}).  In the classical limit,  the spin-helix states satisfy the Landau-Lifshitz equations of motion $\partial_t \vec{S}_i(t)\,{=}\,(\partial H/\partial \vec{S}_i)\,{\times}\,\vec{S}_i(t)\,{=}\,0$ for any wavevector $Q$ and therefore do not decay (this is in fact true for any anisotropy $\Delta$).  Here, $\vec{S}_i$ is a classical spin vector, which corresponds to the $S\,{\to}\,\infty$  limit of a quantum mechanical spin.  For finite $S$ we can study the effects of quantum fluctuations with a large spin ($1/S$) expansion. We find that the Fourier modes of the fluctuations carrying momentum $k$ have a dispersion relation $\omega_k\,{\propto}\,J_{xy}|\cos(Qa) \sin(k/2)|$. As the characteristic energy scales of all modes are proportional to $\cos(Qa)$, this indicates, in a somewhat surprising fashion, that the dynamics of helices with longer wavelengths is faster than for smaller wavelengths, with the slowest dynamics occurring at $Q\,{=}\,\pi/(2a)$. This prediction is furthermore corroborated by a fully quantum ($S\,{=}\,1/2$) but short-time expansion of the order parameter of the spin helix (see Appendix \ref{Appendix:ShortTime}). Numerical simulations, as seen in Fig.~\ref{fig:decayCurves}b and Fig.~\ref{fig:XX_theory}a of the Appendix, also verify this by showing  a very good collapse of the decay curves of all experimentally considered wavevectors $Q$ upon rescaling time by a factor of $\cos(Qa)$. This holds even up to evolution times $t$ longer than would be expected to be valid for the semiclassical analysis or short-time expansion. As expected, deviations from this relation are seen as the wavevector approaches $Q\,{=}\,\pi/(2a)$, for which the simple approaches would predict a vanishing decay rate.  
 
Experimentally, we find that the  decay rates of the transverse helices as a function of wavevector $Q$ can be fitted very well as the sum of the predicted $\cos(Qa)$ dependence together with a constant term, as shown in Fig.~\ref{fig:decayCurves}b.  The constant term represents additional dephasing mechanisms that go beyond the idealizations of the spin model \eqref{Heisenberg_eq}, which we will discuss below.

\begin{figure}[t]
	\includegraphics[width=\linewidth,keepaspectratio]{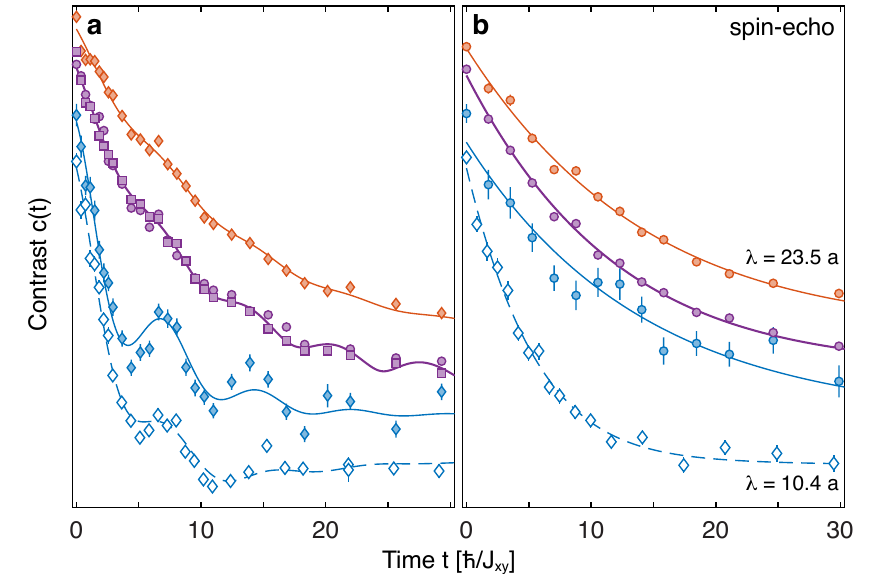}
	\caption{\textbf{Absolute measurement of the effective magnetic field value $\mathbf{h_z}$} as a beat note between the spin precession frequencies in the inner and outer parts of the cloud. \textbf{a}, Transverse spin-helix contrast $c(t)$ for $\lambda\,{=}\,23.5\,a$ (all filled symbols) and $\Delta\,{\approx}\,1$. For measurements integrated over the whole cloud (purple), they were performed at $11\,E_R$ ($\circ$) and $13\,E_R$ ($\Box$). The contrast at the center of the atom cloud for radii $r\,{\leq}\,8.5\,a$ (orange filled symbols) decays slower with less pronounced oscillations, whereas the contrast in the spatial wings for radii $r\,{\geq}\,20\,a$ (blue filled symbols, including ${\approx}\,7\,\%$ of the atoms in the cloud) decays faster with more pronounced oscillations. Data points for orange and blue filled diamonds are an average of measurements at $11\,E_R$ and $13\,E_R$. Open diamond symbols represent data for $\lambda\,{=}\,10.4\,a$ and show the same oscillation frequency but decay faster due to spin transport. Solid lines are fits described in Appendix \ref{Appendix:Analysis}. \textbf{b}, Identical measurements as in panel \textbf{a}, but with a spin-echo pulse which removes the oscillations. Curves in both \textbf{a} and \textbf{b} are offset from each other for clarity. }
	\label{fig:beatNote}
	\vspace{-0pt}
\end{figure}

\subsection{XXX model}
\label{XXX_model}

When $\Delta\,{=}\,1$, we realize the  isotropic Heisenberg spin model  which, aside from the effective magnetic field $h_z$, satisfies  $[H, \sum_i S^\alpha_i ]\,{=}\,0$ $(\alpha\,{=}\,x,y,z)$. The presence of the effective magnetic field term $\sum_{\langle ij \rangle} \frac{h_z}{2} (S^z_i\,{+}\,S^z_j)$  in Eq.~\eqref{Heisenberg_eq}, which we can rewrite as $\sum_i h_{z,i}S^z_i$,  explicitly breaks this spin-rotational symmetry. Now, a uniform (i.e.~$i$-independent) field $h_{z,i}$  can be transformed away by going into an appropriate rotating frame. In such a case,  transverse and longitudinal spin-helix states should show exactly the same dynamical behavior.  However, as the results in Fig.~\ref{fig:decayCurves}d-e show,  there is a dramatic difference with the transverse helices decaying  much faster than the longitudinal helices. In particular,   longitudinal helices exhibit  a diffusive scaling with wavevector $Q$, i.e.~their decay rates obey $\gamma\,{\propto}\,Q^2$ (as shown in \cite{Jepsen2020_SpinTransport}), whereas the transverse helices have an additional $Q$-independent decay rate of $\gamma_0\,{=}\,0.096(10)\,J_{xy}/\hbar$ which dominates the decay for small values of $Q$. Such observations reveal the presence of explicit symmetry-breaking terms which lead to dephasing mechanisms for the transverse helices. In the following, we identify and quantify three mechanisms:  (i) non-uniformities in the effective magnetic field across our setup, (ii) edge effects from a given chain of finite length, and (iii) fluctuations due to mobile holes in the system. 

We note that in the final data analysis, a reevaluation of the scattering lengths showed that our data was actually not  taken exactly at the isotropic point, but at $\Delta\,{=}\,0.93 \pm 0.05$.  This deviation is responsible for a $Q$-independent decay rate of $0.015\,J_{xy}/\hbar$, or $15\,\%$ of the observed difference between longitudinal and transverse spin decay (Appendix \ref{subsection:XXZ_appendix}).

\subsection{Imaging the effective magnetic field}

 First, we  present direct experimental observations of the effective magnetic field. 
If different chains in the ensemble of our setup experience different (real or effective) magnetic fields, then transverse helices precess at different rates, and when the spin pattern is averaged over all the chains, the measured contrast decays. If the ensemble has two pronounced values of the effective magnetic field, then the time evolution of the cloud averaged contrast $c(t)$ shows a beat note at a frequency which corresponds to the difference of the two values of the effective magnetic field.  This is the case for our atom clouds, which feature a Mott insulator plateau surrounded by individual atoms which are pinned to their lattice sites by the gradient of the harmonic trapping potential. Many of these atoms do not have neighbours for spin exchange, and therefore do not feel an effective magnetic field.  The observed beat frequencies $\omega\,{=}\,0.90(1)\,J_{xy}/\hbar$ (Fig. \ref{fig:beatNote}) agree well with the predicted value of the effective magnetic field $h_z\,{=}\,0.89\,J_{xy}$.  As expected, the beat note is more pronounced by spatially selecting the outer parts of the cloud, and disappear with the spin-echo. In Appendix \ref{Appendix:Spectroscopy} we describe an alternate spectroscopic method to observe the effective magnetic field as a frequency shift.

\begin{figure}[t!]
	\includegraphics[width=\linewidth,keepaspectratio]{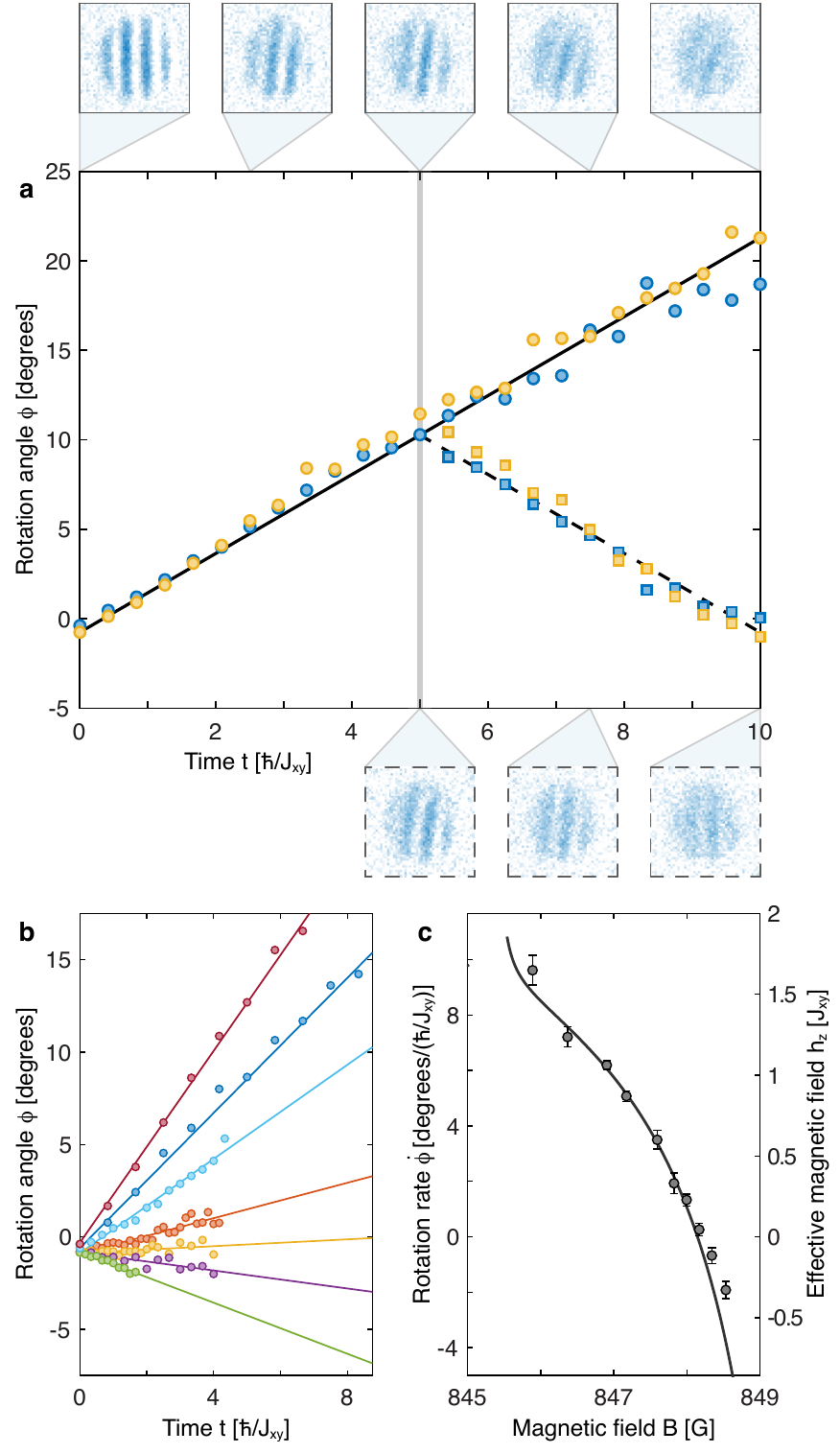}
	\caption{\textbf{Direct observation of the effective magnetic field through spin precession}. \textbf{a}, Rotation angle of the stripe pattern as function of evolution time $t$ (filled symbols) for two lattice depths $11\,E_R$ (blue) and $13\,E_R$ (yellow) without (solid line) and with (dashed line) spin-echo pulse at $t\,{=}\,5\,\hbar/J_{xy}$.  \textbf{b},\textbf{c}, Tunability of the effective magnetic field $h_z$. \textbf{b}, Rotation angle as a function of evolution time $t$ for different magnetic fields fields $B\,{=}\,846.37\,\text{G}$ (red), $847.17\,\text{G}$ (blue), $847.59\,\text{G}$ (light blue), $848.00\,\text{G}$ (orange), $848.17\,\text{G}$ (yellow), $848.34\,\text{G}$ (purple), $848.53\,\text{G}$ (green). \textbf{c}, Angular velocities obtained from linear fits in \textbf{b} compared to predicted effective magnetic fields $h_z$ (solid line) with the scale factor between the two y-axes as a fitting parameter, representing the (uncalibrated) gradient of the lattice depth. Times are normalized by the spin-exchange time $\hbar/J_{xy}$ for the central part of the atom cloud. }
	\label{fig:spinPrecession}
	\vspace{-0pt}
\end{figure}

The presence of the effective magnetic field can also be directly imaged by introducing a sufficiently large gradient in the lattice depth across the chains. This can be achieved by vertically displacing the $z$-lattice relative to the atom cloud (see Fig.~\ref{fig:setup}), causing a gradient of the effective magnetic field. As this field sets  the \enquote{spiraling} frequencies of the helices across the cloud (simply arising from the on-site precession of the spins about the $z$-axis), this translates to an observable tilt of the whole stripe pattern (Fig.~\ref{fig:spinPrecession}a). 

The tilt angle grows linearly in time, with a rate proportional to the effective magnetic field, and the gradient of the lattice depth (which is kept fixed). Externally applied magnetic fields change the scattering lengths via broad Feshbach resonances, and so we can tune the effective magnetic field.   The observed rates for the tilt rotation versus applied magnetic field $B$ are shown in Fig.~\ref{fig:spinPrecession}b-c and agree well with our theoretical prediction.  In particular, near $B\,{=}\,848.1\,\text{G}$, the spin $\ket{\uparrow}$ and $\ket{\downarrow}$ scattering lengths are identical $a_{\uparrow\uparrow}\,{=}\,a_{\downarrow\downarrow}$  and the effective magnetic field is zero, evinced by the absence of any tilt in time (Fig.~\ref{fig:spinPrecession}b; yellow data points).

We note that in principle, such a rotation could also be caused by an external magnetic field gradient. However, the tilt angle would then not depend on lattice depths and external magnetic field $B$.  In a deep lattice, for up to at least $40\,\text{ms}$, we do not observe any discernible rotation, hence ruling out an external field gradient.

When an echo pulse is added, the direction of the stripe rotation is reversed, and at twice the echo time, the stripe pattern is vertical again, resulting in high contrast for vertically integrated images (see Fig.~\ref{fig:spinPrecession}a; bottom images). This shows how the spin-echo eliminates the effect of inhomogeneous effective fields across the cloud, a technique which we will use below to quantify their contribution to the dephasing of transverse spin patterns.

\subsection{Dephasing mechanisms for the transverse spin helix}

In the following we explain how the effective magnetic field leads to different dephasing mechanisms for the transverse helices.

We first investigate the effect of the inhomogeneity in the effective magnetic field strengths across different chains in the ensemble, present due to the slight variations in the lattice depth caused by the Gaussian shape of the laser beams. Such an effect can be eliminated by applying a spin-echo pulse at half of the evolution time $t$. Indeed, upon applying the spin-echo, the $Q$-independent background decay rate $\gamma_0$ is reduced from $0.096(10)$ to $0.060(3)\,J_{xy}/\hbar$ (dashed lines in Fig.~\ref{fig:decayCurves}e). This is compatible with an  effective magnetic field distribution over different chains with a full-width at half-maximum (FWHM) of $8.2\,\%$, corresponding to variations in the lattice depth $V_z$ of $1.6\,\%$, compatible with experimental parameters.

Next, we investigate the effect of   finite chain lengths. We observe that even for  uniform lattice depths, the effective magnetic field is {\it necessarily} non-uniform at   the ends of  a given chain. This has observable consequences which have not been discussed before. To elaborate, note that the effective magnetic field arises from superexchange processes involving nearest neighbor pairs of atoms, indicated already in the Hamiltonian Eq.~\eqref{Heisenberg_eq} where we have deliberately written the magnetic field  $\sum_{\langle i j \rangle}( S^z_i + S^z_j )$ as a sum over pairs $\langle ij \rangle$ of sites to emphasize this fact.  This means that for a 1D chain, the effective magnetic field is reduced to $h_z/2$ at the ends, half the value in the bulk. For $h_z\,{\neq}\,0$, this non-uniformity can hence not simply be transformed away by going in an appropriate co-rotating frame. 

This edge effect results in differences in the relaxation of transverse and longitudinal spin helices:  The spins at the edges dephase rapidly, and this perturbation then propagates through the entire chain. We have performed numerical simulations which show that for chain lengths of $10$ - $20$ spins, the edge effect causes a dephasing rate of ${\approx}\,0.02\,J_{xy}/\hbar$ (Section \ref{sec:dephasing_edge} of the Appendix). Although the diameter of the Mott insulator plateau is around 40 sites, we expect that holes   (with an estimated concentration of $5$ - $10\,\%$) localized by the trapping potential   effectively create such shorter chains in our sample.

\begin{figure}[t]
	\includegraphics[width=\linewidth,keepaspectratio]{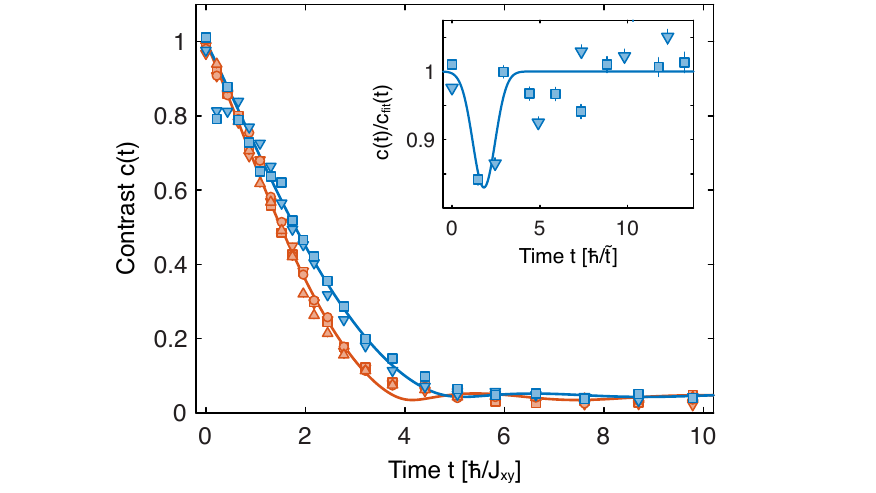}
	\caption{\textbf{Tunneling dynamics of mobile holes}. Decay curves for the XX model ($\Delta\,{\approx}\,0$) at $\lambda\,{=}\,10.4\,a$ without a spin-echo pulse (red) and with a spin-echo pulse (blue) after half the evolution time $t$, measured for different lattice depths $9\,E_R$ ($\bigtriangleup$), $11\,E_R$ ($\circ$), $13\,E_R$ ($\Box$), $15\,E_R$ ($\bigtriangledown$). The time axis has units of the corresponding spin-exchange times $\hbar/J_{xy}\,{=}\,0.64\,\text{ms} $ ($\bigtriangleup$), $1.71\,\text{ms}$ ($\circ$), $4.30\,\text{ms}$ ($\Box$), $10.29\,\text{ms}$ ($\bigtriangledown$). Solid lines are fits (see Appendix \ref{Appendix:Analysis}). The spin-echo generally slows down relaxation slightly by eliminating inhomogeneous dephasing, except at early times where relaxation is actually enhanced and the data points reproducibly deviate from the solid line (data points at $t\,{=}\,0.22$ and $0.43\,\hbar/J_{xy}$ had to be excluded from fitting). The inset shows the data points divided by the fit function and plotted with time in units of the tunneling time $\hbar/\tilde{t}$, showing enhanced relaxation around 1-2 tunneling times (solid line is a guide for the eye). }
	\label{fig:holeDynamics}
	\vspace{-0pt}
\end{figure}

In addition to the effect at the ends of the chain, the harmonic trapping potential along the chains creates an additional inhomogeneity of the effective magnetic field, since the superexchange rate is modified by the offset between neighbouring sites due to the trapping potential by approximately 10 \% (see \cite{Jepsen2020_SpinTransport} for details).  We estimate that this is somewhat less important than the 50 \% effect at the ends of the chain.

We also explore yet another mechanism for how the effective magnetic field can break the symmetry of the isotropic spin Hamiltonian: the presence of mobile holes, not captured by the pure spin model \eqref{Heisenberg_eq}. In a simplified picture, holes play a role  because spins next to holes experience only half the effective magnetic field.  A mobile hole will therefore create a fluctuating effective magnetic field, causing dephasing of the transverse spin component. We will make this picture more  precise by considering the bosonic $\tilde{t}$-$J$ model later.

We have experimentally observed evidence for dephasing by mobile holes, causing a feature in the contrast at early times on the order of the tunneling time $\hbar/\tilde{t}$ (Fig.~\ref{fig:holeDynamics}). 
Let us model the dynamics of holes by a quantum random walk where the  time-dependent  wavefunction at site $i$ for a hole initially localized at $i\,{=}\,0$ is the Bessel function $J_i(t/ (\hbar/2\tilde{t}))$. The square of the Bessel function shows oscillations at frequencies $\omega\,{=}\,4 \tilde{t}/ \hbar$ or periods of $T=(\pi/2) \hbar/\tilde{t}$. For a hold time $T$, with an echo pulse at $T/2$, those fluctuations are \enquote{rectified} and lead to enhanced dephasing (Fig.~\ref{fig:holeDynamics}). This is evidence for hole-magnon coupling:  holes carry a localized magnetic field which couples to spin dynamics.  We have only taken sparse data at very short times, and will address hole dynamics more thoroughly in future work.  By using a series of echo pulses at frequency $\omega$, one could map out the frequency spectrum of the effective magnetic field, using concepts from dynamic decoupling \cite{Cappellaro2017}.
 
The effect of the hole-induced fluctuating effective magnetic field can be captured by a simple model.  From nuclear magnetic resonance, it is well known that the dephasing time $T_2$ of a localized spin at $z\,{=}\,0$ is related to the magnetic field fluctuations $h_z$ (measured in units of energy) and their coherence time $\tau_c$ \cite{Slichter2020} via 
$1/T_2\,{=}\,\langle h_z^2 \rangle \tau_c / \hbar^2\,{=}\,G(z\,{=}\,0,\omega\,{=}\,0) / \hbar^2 $
where $G(z,t)\,{=}\,\langle h_z(z,t) h_z(0,0) \rangle $ is the auto-correlation  of the fluctuating magnetic field along the chain. The dephasing time $T_2$ is the same for spin patterns with arbitrary wave vector.

For a moving hole, the effective magnetic field has a correlation function $G(z,t)$ which is identical to the (normalized) density-density correlation function $J(z,t)$ of the hole, multiplied by $h_z^2$ (here we neglect the fact that the effective magnetic field at a given site depends on the holes on the neighbouring sites, see Eq.~\eqref{Heisenberg_eq}).  For uncorrelated holes with hole probability $p$, the variance of the local occupation is $p$ with a coherence time $\tau_c\,{=}\,1.14 (\hbar/2\tilde{t})$ where $\tilde{t}$ is the tunneling matrix element \cite{footnotePrefactor}.  The associated correlations of the fluctuating effective magnetic field determine the $T_2$ dephasing time for the spin helix $1/T_2\,{=}\,0.57 h_z^2 p /\tilde{t} \hbar$.  Assuming $p\,{=}\,0.1$ ($10\,\%$ hole fraction)  and using $\tilde{t}\,{=}\,6.15\,J_{xy}$ (at $11\,E_R$) and $h_z/J_{xy}\,{=}\,0.89$, this leads to an estimate of $1/T_2\,{\approx}\,0.007\,J_{xy}/\hbar$.  Below, we substantiate this simple model with a full simulation of the bosonic $\tilde{t}$-$J$ model which explicitly takes into account the presence of holes in the Mott insulator near unity filling.

\subsection{Bosonic $\tilde{t}$-$J$ model} 
Our experimental observations show differences to predictions from the pure spin model and provide strong hints for the presence of mobile holes in the underlying Mott insulator  which in typical experiments is on the order of 5 -  10 \% \cite{Bloch2014_SpinHelix, Jepsen2020_SpinTransport}.  We therefore generalize the spin model to the bosonic $\tilde{t}$-$J$ model where holes are present, but double occupancy is suppressed by the large on-site repulsion $U$ (see Appendix \ref{Appendix:tJmodel} for the derivation beginning from the Bose-Hubbard model):
\begin{widetext}
	\vspace{-5pt}
	\begin{align}
		H_{\tilde{t}\text{-}J} & = \sum_{\langle i j \rangle} \left[ J_{xy} (S^x_i S^x_j + S^y_i S^y_j) + J_z S^z_i S^z_j -  \frac{h_z}{2} \left( S^z_i(n_{\uparrow j} +  n_{\downarrow j}) + (n_{\uparrow i } + n_{ \downarrow i}) S^z_j \right)  + c (n_{\uparrow i} + n_{\downarrow i })(n_{\uparrow j} + n_{\downarrow j})  \right]  \nonumber \\
    & - \sum_{\sigma, \langle ij \rangle} \tilde{t} a_{\sigma i}^\dagger a_{\sigma j} + \text{h.c.}   - \sum_{\sigma,\langle ijk\rangle } \left[ \frac{\tilde{t}^2}{U_{\uparrow \downarrow}} a^\dagger_{\sigma i} n_{\bar{\sigma} j} a_{\sigma k} + \frac{\tilde{t}^2}{U_{\uparrow \downarrow}} a_{\bar\sigma i}^\dagger S^\sigma_j a_{\sigma k} + \frac{2\tilde{t}^2}{U_{\sigma \sigma}} a_{\sigma i}^\dagger n_{\sigma j} a_{\sigma k} \right] + \text{h.c.},
		\label{eqn:tJ}
	\end{align}
	\vspace{-5pt}
\end{widetext}
with spin ${\sigma}\,{=}\,{\uparrow,\downarrow}$. 
Here $a_{\sigma i}$, $a^\dagger_{\sigma i}$ are bosonic lowering and raising operators at site $i$.  $S^\uparrow_i\,{\equiv}\, S^+_i\,{:=}\,a^\dagger_{\uparrow i} a_{\downarrow i}$, $S^\downarrow_i\,{\equiv}\, S^-_i\,{=}\,(S^+_i)^\dagger$, $S^x_i\,{:=}\,\frac{1}{2}(S^+_i\,{+}\,S^-_i)$, $S^y\,{:=}\,\frac{1}{2}(-i S^+_i\,{+}\,i S^-_i)$, $S^z\,{:=}\,\frac{1}{2}(n_{\uparrow i}\,{-}\,n_{\downarrow i})$. Also, $c\,{=}\,-\tilde{t}^2(1/U_{\uparrow \uparrow}\,{+}\,1/U_{\downarrow \downarrow}\,{+}\,1/U_{\uparrow \downarrow} )$. We restrict the on-site Hilbert space to be spanned by three states: an occupancy of a single boson (of either spin species) or no bosons (hole). 

The above Hamiltonian illustrates that the previously identified effective magnetic field term $\sum_{\langle ij \rangle} \frac{h_z}{2} (S^z_i\,{+}\,S^z_j)$ in fact stems from a direct {\it interaction} term describing magnon-density coupling $S^z_i(n_{\uparrow j }\,{+}\,n_{\downarrow j})\,{+}\,(i\,{\leftrightarrow}\,j)$, the latter of which reduces to the former upon taking the limit of no holes i.e.~$(n_{\uparrow i }\,{+}\,n_{\downarrow i})\,{=}\,1$ for every site $i$. The terms on the lower line represent dynamics of holes  in different flavors: bare tunneling, density-assisted tunneling, and spin-flip assisted tunneling, which are additional magnon-hole couplings. Note that they arise at the same order in perturbation theory (in $\tilde{t}/U_{\sigma \sigma'}$) as the pure spin-couplings, and thus in principle cannot be neglected, although they {\it are} suppressed by the presence of a small hole probability $p$. By inspecting the expressions of $J_z$, $h_z$, and these extra terms as a function of $U_{\uparrow\downarrow},U_{\downarrow\downarrow},U_{\uparrow\downarrow}$, we see that at $\Delta\,{=}\,1$ (isotropic spin couplings), a non-zero $h_z$ is a {\it necessary} and {\it sufficient} condition for the   $\tilde{t}$-$J$ model to break spin rotational symmetry. (Having spin rotational symmetry would require all $U_{\sigma \sigma'}$ to be equal). This justifies our identification of the effective magnetic field as the agent giving rise to differences in dynamics between the transverse and longitudinal helices.

The $\tilde{t}$-$J$ model in Eq.~(\ref{eqn:tJ}) was used for both the XX and XXX cases to quantify the rate of dephasing by mobile holes.
Numerical simulations for a mobile hole fraction of $10\,\%$ show additional dephasing at a rate around $0.026\,J_{xy}/\hbar$ for the experimental conditions  in Fig.~\ref{fig:decayCurves}e (see Appendix \ref{subsection:XXX_appendix}), supporting the simplified model of hole-induced dephasing presented above (but providing a rate three times higher than the simple model).  {We have now fully accounted for the isotropy breaking dephasing rate of 0.096 (in units of $J_{xy}/\hbar$) through spin echo (0.036), small deviation from isotropy (0.015), edge effect (0.020)  and mobile holes (0.013-0.026, assuming a hole fraction $p$ between 5 and 10 \%, and linear dependence on $p$).  We regard some of the numbers as only semi-quantitaive due to the non-exponential character of the measured and calculated decay curves.}

The hole-induced dephasing mechanisms observed for the XXX model are also present in the XX model. There the $h_z\,{=}\,1.43\,J_{xy}$ term is even $1.6$ times larger, accounting for the observed $Q$-independent dephasing rate $\gamma_0$ in addition to the $\cos(Qa)$-dependence (Fig.~\ref{fig:decayCurves}b). The amplitude $\gamma_1$ of the $\cos(Qa)$-dependence is a function of the concentration of holes (see Fig.~\ref{fig:decayCurves}b and Fig.~\ref{fig:XX_theory}). We find that the experimental data agree best with numerical simulations for $5\,\%$ holes.

\section{Conclusions}

In conclusion, we have used ultracold atoms to implement the Heisenberg model with tunable anisotropy.  For the relaxation of transverse spin patterns, we have studied for the first time four decay mechanisms: intrinsic dephasing by anisotropic spin exchange couplings, inhomogeneous dephasing through a static superexchange induced effective magnetic field, dephasing through the ends of the chain, and  dephasing by a fluctuating effective magnetic field due to the presence of mobile holes. One reason why several of these mechanisms have not been observed before is that  all previous studies of spin transport in optical lattices have either used fermions \cite{Zwierlein2019_SpinTransport} for which the $\tilde{t}$-$J$ model is always explicitly spin rotationally symmetric  and therefore $h_z\,{=}\,0$, or bosons comprised of $^{87}$Rb \cite{Brown2015_Superexchange2D, Bloch2013_SingleSpin, Bloch2013_BoundMagnons, Bloch2014_SpinHelix} for which the spin $\ket{\uparrow}$ and spin $\ket{\downarrow}$ scattering length are almost identical ($a_{\uparrow\uparrow}\,{=}\,99.0\,a_0$, $a_{\uparrow\downarrow}\,{=}\,99.0\,a_0$ and $a_{\downarrow\downarrow}\,{=}\,100.4\,a_0$) \cite{Kokkelmans2009_rubidiumScatteringLengths}, leading to values of $h_z\,{\approx}\,0.014\,J_{xy}$, approximately 100 times smaller than for $^7$Li.

The experimental and theoretical results presented in this work go beyond pure spin physics.  They illustrate effects caused by a small hole fraction that is generally present in cold atomic quantum simulators.  A more complete description of spin dynamics in such systems  therefore requires using the $\tilde{t}$-$J$ model, which features magnon-hole couplings.

This coupling between density and spin is analogous to the interplay of spin and charge degrees of freedom in strongly-correlated electronic systems, which is important, for example in understanding emergent many-body phenomena like high-temperature superconductivity in the cuprates.  Therefore, our platform presents an elegant new setting where such physics can be emulated. More generally, we regard our work as a starting point for exploring spin dynamics in different dynamical regimes as well as in generalized Heisenberg models. Exciting future directions include the study of spin polaron dynamics \cite{PhysRevB.44.317}, realizing long-lived, metastable prethermal states in higher-dimensions \cite{Knap2015, PhysRevLett.125.230601, rodrigueznieva2020transverse}, and probing the onset of  turbulent spin relaxation utilizing larger spin quantum numbers \cite{rodrigueznieva2020turbulent}.

\textbf{Acknowledgements.} We thank Y.~K.~Lee for experimental assistance, J.~Rodriguez-Nieva and M.~D.~Lukin for discussions, and J.~de Hond for comments on the manuscript. We acknowledge support from the NSF through the Center for Ultracold Atoms and Grant No. 1506369, ARO-MURI Non-Equilibrium Many-Body Dynamics (Grant No.~W911NF-14-1-0003), AFOSR-MURI Photonic Quantum Matter (Grant No.~FA9550-16-1-0323),  AFOSR-MURI Quantum Phases of Matter (Grant No.~FA9550-14-1-0035),  ONR (Grant No.~N00014-17-1-2253), the Vannevar-Bush Faculty Fellowship, and the Gordon and Betty Moore Foundation EPiQS Initiative (Grant No.~GBMF4306). W.W.H.~is supported in part by the Stanford Institute of Theoretical Physics. Numerical simulations involving matrix product states were performed using the TeNPy Library \cite{tenpy}.

\clearpage
 
\appendix

\onecolumngrid

\begin{figure}[t]
	\includegraphics[width=\linewidth,keepaspectratio]{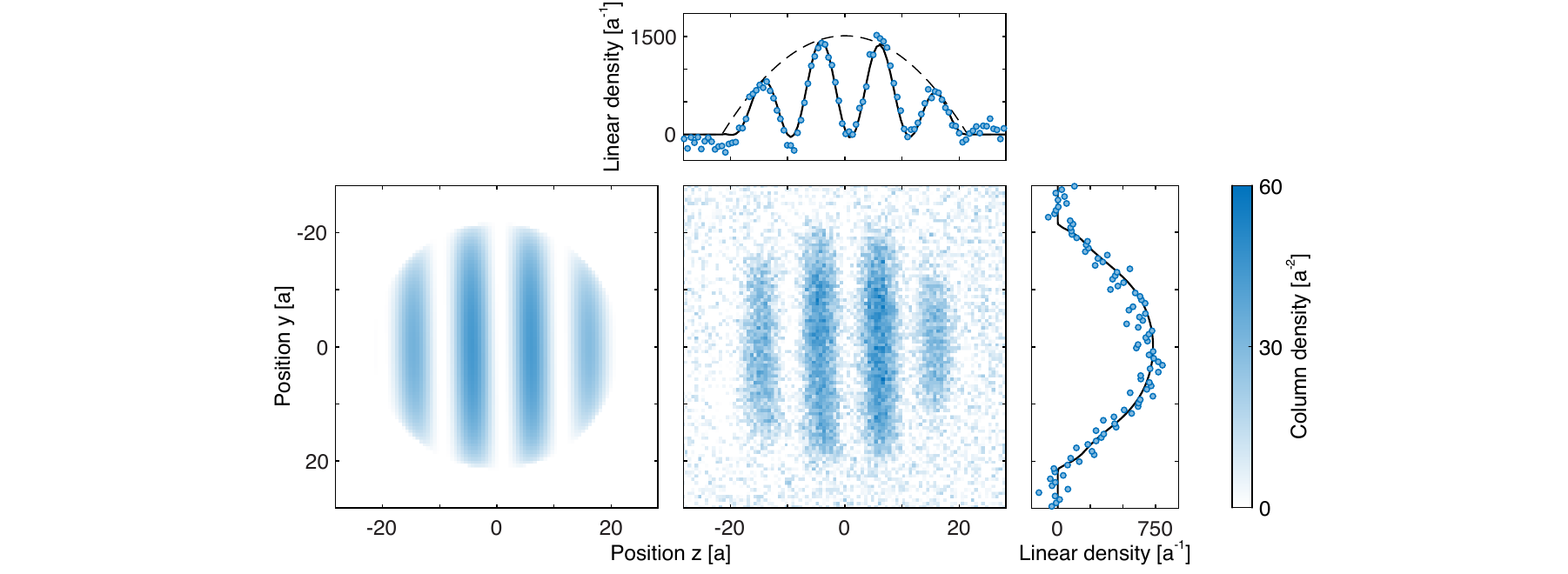}
	\caption{\textbf{Contrast measurement}. The central image represents raw data and shows the distribution of atoms in the $\ket{\uparrow}$ state. The left panel is the two-dimensional fit as described in Appendix \ref{Appendix:Analysis}. Every pixel is a local measurement of the column density (number of $\ket{\uparrow}$ atoms per unit area). The image is projected (integrated) both along the horizontal ($z$) and vertical ($y$) direction from $x$,$y\,{=}\,{-}42\,a$ to ${+}42\,a$ to obtain the linear densities (number of $\ket{\uparrow}$ atoms per unit length) for the measured data (points) and the fit (solid line). The dashed line in the $y$-projection shows the (parabolic) envelope function for the spatial density distribution in the cloud. The $y$- and $z$-axis are displayed in units of the lattice spacing $a\,{=}\,0.532\,\mu\text{m}$. }
	\label{fig:2Dfit}
	\vspace{-9pt}
\end{figure}

\section{Data analysis}
\label{Appendix:Analysis}

We determine the contrast $\mathcal{C}$ by a fit $f(y,z)\,{=}\,{g(y,z)\,{\cdot}\,[{1}\,{+}\,{\mathcal{C} \cos(Qz+\theta)}]/2}$ to the two-dimensinal phase-contrast images.  Here $Q\,{=}\,2\pi/\lambda$ is the wavevector, $g(y,z)$ is a two-dimensional envelope function which accounts for the spatial distribution of all atoms $n\,{=}\,n_\uparrow\,{+}\,n_\downarrow$ inside a sphere of radius $R$ such that $g(y,z)\,{=}\,A\,[R^2-y^2-z^2]^{1/2}\,{\cdot}\,H(R^2-y^2-z^2)$ with the Heaviside function $H(r)$, and $\theta$ is a random phase which varies from shot to shot due to small magnetic bias field drifts. During the evolution time $t$ the contrast $\mathcal{C}(t)$ decays, and we study the dependence of $c(t)\,{=}\,\mathcal{C}(t)/\mathcal{C}(0)$ on lattice depth $V_z$, wavelength $\lambda$, and anisotropy $\Delta$.

For both longitudinal and transverse spin relaxation, we find the decay curves can be well described by the sum of a decaying part with time constant $\tau$ and a (damped) oscillating part with frequency $\omega$, resulting in a fitting function $c(t)\,{=}\,{\left( a_0\,{+}\,b_0\cos\omega t\right) e^{-t/\tau}}\,{+}\,c_0$. Here $a_0$, $b_0$, $c_0$, $\omega$, $\tau$ are fitting parameters. These fits were used for Fig.~\ref{fig:decayCurves}b (purple), in Fig.~\ref{fig:decayCurves}d (blue), and for Fig.~\ref{fig:decayCurves}e (purple and blue). Special fitting procedures were used for the XX model and for the beat frequency due to the effective magnetic field.

{\it XX model}:  The decay curves $c(t)$ for the transverse spin helix (Fig.~\ref{fig:decayCurves}a) clearly show a slower decay rate for larger $Q$. We could use the fitting function described above, with the only difference of adding the constant offset $c_0$ in quadrature $c(t)\,{=}\, \sqrt{ {[\left( a_0\,{+}\,b_0\cos\omega t\right) e^{-t/\tau}}]^2\,{+}\,c_0^2 }$ (Fig.~\ref{fig:decayCurves}a) to reflect that the offset $c_0$ arises due to an experimental detection noise floor at the $10^{-2}$. The actual physical contrast does decay to zero $c(t)\,{\to}\,0$. Remarkably, the fitted oscillation periods (e.g. $T\,{=}\,11.6(4)\,\hbar/J_{xy} $ at $\lambda\,{=}\,10.4\,a$) agree fairly well with the energy splitting $J_{xy}/2$ for the two-spin Heisenberg model (Fig.~\ref{fig:decayCurves}c) implying an oscillation period of $T\,{=}\,4\pi\,\hbar/J_{xy}\,{\approx}\,12.57\,\hbar/J_{xy}$. With this fit function, the slower decay for large wavevectors $Q$ shows up mainly in the oscillation frequency $\omega$ and not in the decay time $\tau$. For a simpler characterization of the decay, we obtain the initial decay rate by fitting a linear slope $c(t)\,{=}\,c_0 (1-t/\tau)$ to the initial decay in the range $c(t)\,{\leq}\,0.4$. Results of such fits are shown in Fig.~\ref{fig:decayCurves}b (red).

{\it XXX model}: {\it (1) Longitudinal spin relaxation} (Fig.~\ref{fig:decayCurves}e, purple): As in our previous work \cite{Jepsen2020_SpinTransport} on spin transport, we use the fitting function $c(t)\,{=}\,{\left( a_0\,{+}\,b_0\cos\omega t\right) e^{-t/\tau}}\,{+}\,c_0$. {\it (2) Transverse spin relaxation with spin-echo} (Fig.~\ref{fig:decayCurves}d-e, blue): The same fitting function yields oscillation frequencies $\omega$ and oscillating fractions $b_0/(a_0+b_0)$ which agree fairly well with the longitudinal case, especially at large wavevectors $Q$, but with much larger error bars at small wavevectors $Q$, because the decay time $\tau$ is much shorter. For this reason, we constrain both parameters to the values obtained in the longitudinal case. {\it (3) Transverse spin relaxation without spin-echo} (Fig.~\ref{fig:decayCurves}d-e, red; Fig.~\ref{fig:beatNote}): A beat note between the inner part and outer part of the cloud is visible (Fig. \ref{fig:beatNote}), due to the difference in effective magnetic fields. To determine the beat frequency $\Omega$, we generalize the fitting function to the sum of two parts: $| c_1(t)\,e^{i\Omega t}\,{+}\,c_2(t)|\,{=}\,\sqrt{c_1(t)^2\,{+}\,2c_1(t)c_2(t)\cos(\Omega t)\,{+}\,c_2(t)^2}$.  Here, $c_1(t)\,{=}\,{\left( a_0\,{+}\,b_0\cos\omega t\right) e^{-t/\tau}}$ is the contrast of the atoms in the inner part of the cloud, and $c_2(t)\,{=}\,c_2$ is the contrast of the isolated atoms in the outer part which preserve the contrast for a long time.  We can neglect the background $c_0$ due to the detection noise.  In $c_1(t)$ we again constrain the two parameters $\omega$ and $b_0/(a_0+b_0)$ to the values obtained for longitudinal spin relaxation.

\section{Spectroscopic observation of the effective magnetic field}
\label{Appendix:Spectroscopy}
\begin{figure}[t]
	\includegraphics[width=89mm,keepaspectratio]{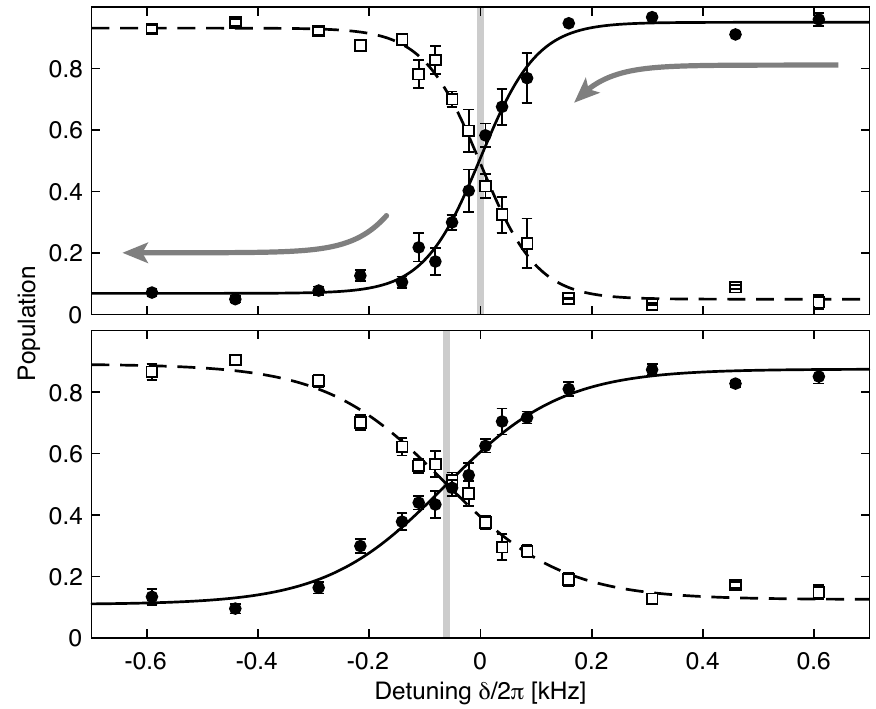}
	\caption{\textbf{Spectroscopic observation of the effective magnetic field $\mathbf{h_z}$}. Shown is the fraction of atoms in each state as a function of the final detuning $\delta$ of a $22\,\text{ms}$ sweep of the RF frequency, starting at $\delta\,{=}\,{+}30\,\text{kHz}$ with all atoms in the $\ket{\uparrow}$ state (closed circles) and no atoms in the $\ket{\downarrow}$ state (open squares). The detuning is relative to the single-particle transition frequency. The power of the RF drive is also ramped to zero after the frequency sweep to make the transition sharper.  A non-zero detuning $\delta$ for equal spin populations compensates for the effective magnetic field $h_z$ which shifts the curves for $11\,E_R$ (bottom panel) lattice depth compared to $35\,E_R$ (top panel) where $h_z\,{\approx}\,0$.  For the sweep experiment, we chose the second-lowest (closed circles) and third-lowest (open squares) hyperfines states of $^7$Li due to the smaller sensitivity to external magnetic fields. At $1025\,\text{G}$, the anisotropy is $\Delta\,{=}\,{-}0.24$. }
	\label{fig:effectiveField}
	\vspace{-0pt}
\end{figure}

For constant density, the constant effective field can always be transformed away in a suitable rotating frame.  However, it can still be observed as a shift of the spin-flip resonance.   We rotate the spins via an adiabatic sweep of frequency, where the detuning corresponds to an external $z$-field (in the rotating frame), and the Rabi frequency to an $x$-field, realizing a Heisenberg model with magnetic fields. Starting from a fully polarized state with all atoms in the $|{\downarrow}\rangle$ state and large detuning of the RF, we reduce  the $z$-field adiabatically and observe the spin imbalance.  The spins are balanced when the detuning compensates for the effective magnetic field created by the superexchange. With this method, we can observe the effective field for different lattice depths (Fig.~\ref{fig:effectiveField}).

\section{Semiclassical analysis and short-time expansion of spin dynamics} 

\subsection{Classical transverse helices of any $Q$ do not evolve for any anisotropy $\Delta$}
\label{Appendix:Semiclassical} 

We show here that in the classical limit,  transverse helices of any wavevector $Q$ do not evolve under the XXZ Hamiltonian (assuming the effective magnetic field is uniform), for any isotropy $\Delta$. {The classical limit is reached by taking the spin-quantum number $S \to \infty$, or by treating the spins as classical vectors $\vec{S}_i = (S^x_i, S^y_i, S^z_i)$ of arbitrary length $S = |\vec{S}_i|$ which we set to 1/2 for comparison to the quantum spin system.}

We start with the system initialized at $t = 0$ in the helix state $ S^\pm_i(0) = S e^{\pm i (Q z_i + \theta)},  S^z_i(0) = 0$, and we \enquote{untwist} the helix by  using the rotation
\begin{align}
    \begin{pmatrix}
    S^x_i \\ S^y_i \\ S^z_i 
    \end{pmatrix} =
    \begin{pmatrix}
    -\sin(Q z_i) & 0 & \cos(Q z_i) \\ 
    \cos(Q z_i) & 0 & \sin(Q z_i) \\
    0 & 1 & 0 
    \end{pmatrix}
    \begin{pmatrix}
    T^x_i \\ T^y_i \\ T^z_i
    \end{pmatrix}
\end{align}
(we ignore the phase $\theta$ for simplicity) which gives $\vec{S}_i \mapsto \vec{T}_i$ with $T^\pm_i(0) =0, T^z_i(0) = S$. The Hamiltonian Eq.~\eqref{Heisenberg_eq} (with $h_z = 0$) then transforms as
\begin{align}
    H \mapsto H(Q) = J_{xy} \sum_i \left[ \cos(Qa) (T^x_i T^x_{i+1} + T^z_i T^z_{i+1}) + \sin(Qa)(T^z_i T^x_{i+1}- T^x_i T^z_{i+1}) \right] + J_z \sum_i T^y_i T^y_{i+1}.
\end{align}
The Landau-Lifshitz (LL) equations of motion  for classical spins read $\partial_t \vec{T}_i = \partial_{\vec{T}_i} H(Q) \times \vec{T}_i$. Upon changing variables to $T^\pm_i = T^x_i \pm i T^y_i$ we have
\begin{align}
    \dot{T^+_i} & = \frac{1}{2} i [J_z (T^-_{i-1} - T^-_{i+1} - T^+_{i-1} - T^+_{i+1} )T^z_i \nonumber \\
    & +  J_{xy}(\sin(Qa)T^+_i(T^-_{i-1} - T^-_{i+1} + T^+_{i-1} - T^+_{i+1}) + 2 T^z_i ( T^z_{i-1} - T^z_{i+1} ) ) \nonumber \\
    & + \cos(Qa) (- (T^-_{i-1} + T^-_{i+1} + T^+_{i-1} + T^+_{i+1})T^z_i + 2T^+_i(T^z_{i-1}  + T^z_{i+1})))
    ] \\
    \dot{T^-_i} & = \dot{T^+_i}^* \\
    \dot{T^z_i}& = -\frac{1}{4}i (J_z (T^-_i + T^+_i)(T^-_{i-1} + T^-_{i+1} - T^+_{i-1} + T^+_{i+1}) \nonumber \\
    & - J_{xy}(T^-_i - T^+_i)(\cos(Qa)T^-_{i-1} + T^-_{i+1} + T^+_{i-1} + T^+_{i+1}) + 2 \sin(Qa)(-T^z_{i-1} + T^z_{i+1})))
\end{align}
and it is straightforward to verify  that $T^{\pm}_i(t) = 0, T^{z}_i(t) = S$   is a solution to the LL equations, as claimed.

\subsection{Stability of the classical spin-helix state: dispersion relation of fluctuations}
\label{appendix:dispersion}

To understand the stability of the classical spin-helix states we linearize the equations of motion about the classical solution and consider fluctuations.
Now, $T^z_i$ obeys the constraint $T^z_i = \sqrt{S^2 - T_i^+ T_i^-}$ so   fluctuations about the classical solution simply entails
\begin{align}
    & T^\pm_i(t) = \bar{T}^{\pm}_i(t) + \delta T_i^\pm(t) + O((\delta T)^2)= \delta T_i^\pm(t) + O((\delta T)^2), \\ 
    & T^z_i(t) = S + O((\delta T)^2).
\end{align}
Therefore we get  
\begin{align}
\delta T^\pm_i = \frac{i S}{2} (J_z (\delta T^-_{i-1} + \delta T^-_{i+1} - \delta T^+_{i-1} - \delta T^+_{i+1} ) \mp \cos(Qa) J_{xy} ( \delta T^-_{i-1} + \delta T^-_{i+1} + \delta T^+_{i-1} + \delta T^+_{i+1} - 4 \delta T^\pm_i ))
\end{align}

We now expand in Fourier modes $\delta T^\pm_i = \sum_k \delta T^\pm_k e^{i (k z_i + \omega_k t)}$ with momentum $k$ and dispersion $\omega_k$. This reduces to an eigenvalue problem

\begin{align}
\omega_k
    \begin{pmatrix}
    {\delta T^+_k} \\
    {\delta T^-_k}
    \end{pmatrix} = S J_{xy}
    \begin{pmatrix}
-(-2+\cos(ka))\cos(Qa) - \Delta \cos(ka) & \cos(ka) (-\cos(Qa) + \Delta) \\
\cos(ka)(\cos(Qa) - \Delta) & (-2 + \cos(ka))\cos(Qa) + \Delta \cos(ka) 
\end{pmatrix}
\begin{pmatrix}
    {\delta T^+_k} \\
    {\delta T^-_k}
    \end{pmatrix}
\end{align}
with solution
\begin{align}
     \omega_k = \pm 2 \sqrt{2} J_{xy} S \sqrt{\cos(Qa) (-\Delta \cos(ka) + \cos(Qa))\sin^2(ka/2)}.
    \label{eqn:omega}
\end{align}
Reducing to $\Delta = 0$ reproduces the expression quoted in the main text for the XX model in Sec.~\ref{sec:XXmodel}. Note that we could have equivalently obtained the same dispersion relations by performing a spin-wave analysis, upon mapping the spins to Holstein-Primakoff bosons (in a large $S$ expansion) and peforming a Bogoliubov transformation to diagonalize the Hamiltonian in second order. 

\subsection{Short-time expansion of quantum dynamics}
\label{Appendix:ShortTime}

Owing to the factorizable nature of the initial spin helix state we can analytically derive the short-time quantum dynamics of the state without passing into a semiclassical limit as was done before. The basic object is the Taylor expansion of a spin operator (in the transverse direction), 
\begin{align} 
    \langle S^+_i(t) \rangle = \langle S^+_i(0) \rangle + \langle \partial_t S^+_i(0)\rangle t + \frac{1}{2} \langle \partial_t^2 S^+_i(0) \rangle t^2 + \cdots 
\end{align}
where $\langle \cdot\rangle$ is the expectation value in the spin  state $|\psi(Q)\rangle = e^{-i \sum_i S^z_i  Q z_i    } |+++\cdots\rangle$  ($S^x_i |+\rangle_i = S |+\rangle_i$), and 
\begin{align}
    & \langle \partial_t S^+_i(0)\rangle = i \langle [H, S^+_i] \rangle, \\
    & \langle \partial_t^2 S^+_i(0) \rangle = - \langle [H,[H,S^+_i]] \rangle.
\end{align}
Since the Hamiltonian is a sum of strictly local terms and $S^+_i$ is an on-site term, the expressions in the commutators are only comprised of finite range terms with support centered around site $i$. Using that the state factorizes into a product state we can easily evaluate the expression for these terms. We find for general spin $S$
\begin{align}
     i\langle [H,S^+_i] \rangle & = 0, \\
     - \langle [H,[H,S^+_i]] \rangle &= -S^2 e^{i Q z_i} (J_z - J_{xy} \cos(Qa))^2.
\end{align}
(The vanishing of the term linear in $t$ follows from time-reversal symmetry).
Therefore
\begin{align}
    \langle S^+_i(t)\rangle = S e^{i Q z_i} - \frac{1}{2} S^2 e^{i Q z_i} (J_z - J_{xy} \cos(Qa) )^2 t^2 + \cdots.
\end{align}
Extracting the Fourier component with wavevector $Q$ gives the normalized contrast  
\begin{align}
    c(t) = 1 - \frac{1}{2} S (J_z-J_{xy} \cos(Qa))^2 t^2 + \cdots.
\end{align}
A characteristic energy rate $\gamma$ for the initial quadratic decay can therefore be defined as $c(t) = 1 - \gamma^2 t^2 + \cdots$ yielding
\begin{align}
\gamma := \sqrt{\frac{S}{2}} |J_{xy}(\Delta - \cos(Qa) )|.
\end{align}
Focusing now on $\Delta =0$ and $S = 1/2$, this shows that a helix of wavevector $Q$ decays with a rate going as $\gamma  \propto \cos(Qa)$. For $\Delta = 1$ we recover that $\gamma \propto Q^2$. More generally, for arbitrary anisotropy $\Delta$, there is a critical wavevector $(Q_c a) = \arccos(\Delta)$ where decay is expected to be very slow; seeing such a dependence in experiments would be interesting.

\section{Numerical simulation details}
\label{Appendix:Numerics}
Here we present general details on the numerical simulations performed. Unless otherwise specified, we employed the time-evolving block decimation (TEBD) algorithm on matrix product states (MPS) defined on 1D chains of length $L = 40$, with large enough bond dimension to ensure convergence of local observables to a tolerance $10^{-4}$, via the TeNPy library  \cite{tenpy}. 

In the absence of holes, we simulate the Heisenberg model \eqref{Heisenberg_eq}, with the initial state  the spin helix with wave vector $Q$, which reads locally $|\psi_i(Q)\rangle =  e^{i (Q z_i + \theta) S^z_i} |+\rangle_i$ where $S^x_i |+\rangle_i = |+\rangle_i = \frac{1}{\sqrt{2}} ( |\uparrow\rangle_i + |\downarrow\rangle_i )$. Here $\theta$ is the global phase.
We measure $\langle S^x_i(t)\rangle$ and fit for each time-slice a sine function in space with wavevector $Q$ (allowing the phase to be an independent parameter); the amplitude of the sinusoidal modulation is the numerically determined contrast $c(t)$ normalized to unity at $t = 0$. We also average $c(t)$ over $\theta = 0, \pi/2$ to account for the fact that the global phase of the initial state in the experiments shifts from measurement to measurement (we find that averaging over these two values suffices to reproduce the full averaging over $\theta$).

In the presence of holes we simulate the bosonic $\tilde{t}$-$J$ model \eqref{eqn:tJ}. We assume holes  occur independently on each site with probability $p$. In order to perform ensemble averaging over the different hole positions of the initial state, we employed the following computational trick.
Let the on-site Hilbert space be spanned by the states $|0\rangle$ (vacuum; no holes), $|\uparrow\rangle_i := a_{\uparrow,i}^\dagger |0\rangle$ and $|\downarrow \rangle_i := a_{\downarrow, i}^\dagger |0\rangle$.
We define  a pure state on each site as
\begin{align}
    |\Psi_i(Q)\rangle = e^{i \varphi_i} \sqrt{1-p} |\psi_i(Q)\rangle + \sqrt{p} |0\rangle_i,
\end{align}
where $\varphi_i$ is some phase with value in $[0, 2\pi)$. Clearly, in the limit $p \to 0$, the state $|\Psi(Q)\rangle := \prod_i |\Psi_i (Q)\rangle$ reduces to the pure-spin helix (i.e.~without holes) with wave vector $Q$. Consider now the outer-product of $|\Psi_i(Q)\rangle$ with itself when $p \neq 0$:
\begin{align}
    \rho_i = (1-p) |\psi_i(Q)\rangle \langle \psi_i(Q) | + e^{i \varphi_i} \sqrt{1-p} |\psi_i(Q)\rangle \langle 0|_i + \text{h.c.} + p |0\rangle_i \langle 0|_i.
\end{align}
If we now average $\varphi$ over the interval $[0,2\pi)$ uniformly, the ensemble-averaged reduced density matrix is
\begin{align}
    \bar{\rho}_i = \frac{1}{2\pi} \int_0^{2\pi} d\varphi \rho_i = (1-p) |\psi_i(Q)\rangle \langle \psi_i(Q)| + p |0\rangle_i \langle 0|_i.
\end{align}
This   reproduces the situation where holes occur locally and independently on each site $i$ with probability $p$.

In our simulations we choose a random set of phase angles $(\varphi_1, \cdots, \varphi_N) \in [0,2\pi)^N$, and time-evolve the globally pure state $\prod_i |\Psi_i(Q)\rangle$ under the $\tilde{t}$-$J$ Hamiltonian. We repeat the simulation with different sets of phase angles sampled randomly uniformly in $[0,2\pi)^N$, and then average the extracted contrast.

We find that in practice, there are remarkably only very small variations between different choices of phase angles (i.e.~a given  random choice of $\prod_i |\Psi_i(Q)\rangle$ is a typical configuration), allowing us to perform the ensemble average with relatively few repetitions (at most 50 runs for each $Q$ and global phase $\theta$). We also average over global phases $\theta = 0, \pi$.    

\begin{figure*}[t]
	\includegraphics[width=\linewidth,keepaspectratio]{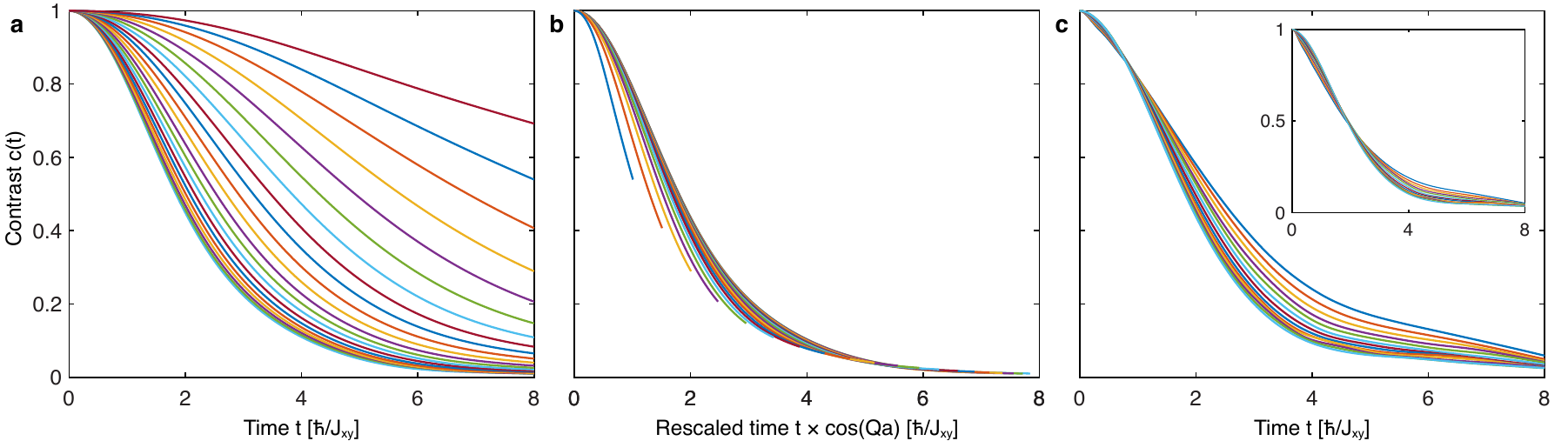}
	\caption{\textbf{Transverse spin-helix decay for the XX model} obtained from numerical simulations, with hole probability $p\,{=}\,0$ (\textbf{a},\textbf{b}), $p\,{=}\,0.05$ (\textbf{c}), $p\,{=}\,0.1$ (\textbf{c}, inset) with $h_z\,{=}\,0$ (\textbf{a},\textbf{b}) and $h_z\,{=}\,1.43\,J_{xy}$ (\textbf{c}) as in the experiment. The wavelengths are (from bottom to top) $\lambda\,{=}\,31.33\,a$, $23.50\,a$, $18.80\,a$, $15.66\,a$, $13.43\,a$, $11.75\,a$, $10.44\,a$, $9.40\,a$, $8.54\,a$, $7.83\,a$, $7.23\,a$, $6.71\,a$, $6.27\,a$, $5.90\,a$, $5.57\,a$, $5.27\,a$, $5.00\,a$, $4.77\,a$, $4.55\,a$, $4.35\,a$, $4.00\,a$. The decay curves for pure spin dynamics (\textbf{a}) show a wavevector dependence of decay rates of the form $\gamma\propto\cos(Qa)$. Using rescaled time $t \cos(Qa)$ all decay curves collapse almost perfectly for wavelengths $\lambda\,{\geq}\,6.27\,a$, which covers the range studied in the experiment. The presence of holes washes out the $\cos(Qa)$ dependence (\textbf{c}), shown here for $\lambda\,{\geq}\,6.27\,a$. At sufficiently high hole probability, e.g. $p\,{=}\,0.1$ (\textbf{c}, inset), the wavevector dependence has vanished almost completely. }
	\label{fig:XX_theory}
	\vspace{-0pt}
\end{figure*} 

\section{Numerical simulation results}

\subsection{XX model}
\label{subsection:XX_appendix}
For the XX model, we utilized parameters $\Delta = 0$, $h_z = 1.43 J_{xy}$ (thus, simulating the experimental conditions at $11E_R$), as well as $\tilde{t} = 4.11 J_{xy}$, $J_{\uparrow \uparrow}/J_{\uparrow \downarrow} = -0.32$, $J_{\downarrow \downarrow}/J_{\uparrow \downarrow} = 1.15$. Here $J_{\sigma \sigma'}:= -4\tilde{t}^2/U_{\sigma \sigma'}$. 

Figure~\ref{fig:XX_theory} shows decay curves for $p\,{=}\,0$, $0.05$, $0.1$. The experimental decay rates $\gamma\,{=}\,1/\tau$ (in Fig.~2b) were obtained from a linear fit $c(t)\,{=}\,c_0 (1\,{-}\,t/\tau)$ to the initial decay ($c(t)\,{\leq}\,0.4$).  We use an equivalent procedure for the theoretical data, and determined the time $\tau'$ where the numerical simulations showed a contrast of $c(\tau')\,{=}\,0.4$, which we then converted to the theoretical decay rate $\tau\,{=}\,(5/3)\tau'$. As shown in Fig.~2b, the numerical data verify the analytically predicted $\cos(Qa)$ dependence of the decay rate. This scaling starts to break down as $Q\,{\to}\,\pi/(2a)$, because higher order terms in the expansion (either semiclassical or short-time) become important. Inclusion of holes washes out the $\cos(Qa)$-dependence.
\\

\subsection{XXX model}
\label{subsection:XXX_appendix}
 
For the XXX model, we utilized parameters $\Delta = 1$, $h_z = 0.80 J_{xy}$ (thus, simulating the experimental conditions at $11E_R$), as well as $\tilde{t} = 6.13 J_{xy}$, $J_{\uparrow \uparrow}/J_{\uparrow \downarrow} = 0.61$, $J_{\downarrow \downarrow}/J_{\uparrow \downarrow} = 1.40$.
 
Figure~\ref{fig:XXZ_theory}a-b show the results for a simulation involving $10\%$ holes, i.e.~$p=0.1$.  We fit the data between $c(t) = 0.9$ to $0.15$ to a straight line, and extract a decay rate defined to be twice  the slope of the fit (this factor is chosen because the slope of an exponential function at half decay is reduced by a factor of two). Due to the non-exponential nature of the decay curves, for both numerical simulations and   experimental data, there is some arbitrariness in choosing \enquote{effective} decay time-constants which, depending on the parametrization, could differ by up to 50 \%. By further fitting the decay rates for the six smallest $Q$ values to the form $\gamma(Q)\,{=}\,DQ^2\,{+}\,\gamma_0$ (in order to focus on the limiting $Q \to 0$ behavior), we obtain a $Q$-independent decay rate $\gamma_0\,{=}\,0.019\,J_{xy}/\hbar$ and a \enquote{diffusion constant} $D\,{=}\,0.16\,a^2/(\hbar/J_{xy})$.
   
 \subsection{Near-isotropic XXZ model }
 \label{subsection:XXZ_appendix}
 
We also investigate how a small deviation from the isotropic point affects dynamics. One reason is that our experimental data for the isotropic model was actually not taken exactly at the isotropic point, but at $\Delta\,{=}\,0.93\,{\pm}\,\,0.05$. In the two-site model, the triplet splitting of $J_{xy}\cdot (1\,{-}\,\Delta)/2$ becomes $0.035 J/ \hbar$, but the full simulation presented here shows that the effect is smaller.
 
\begin{figure*}[t]
	\includegraphics[width=\linewidth,keepaspectratio]{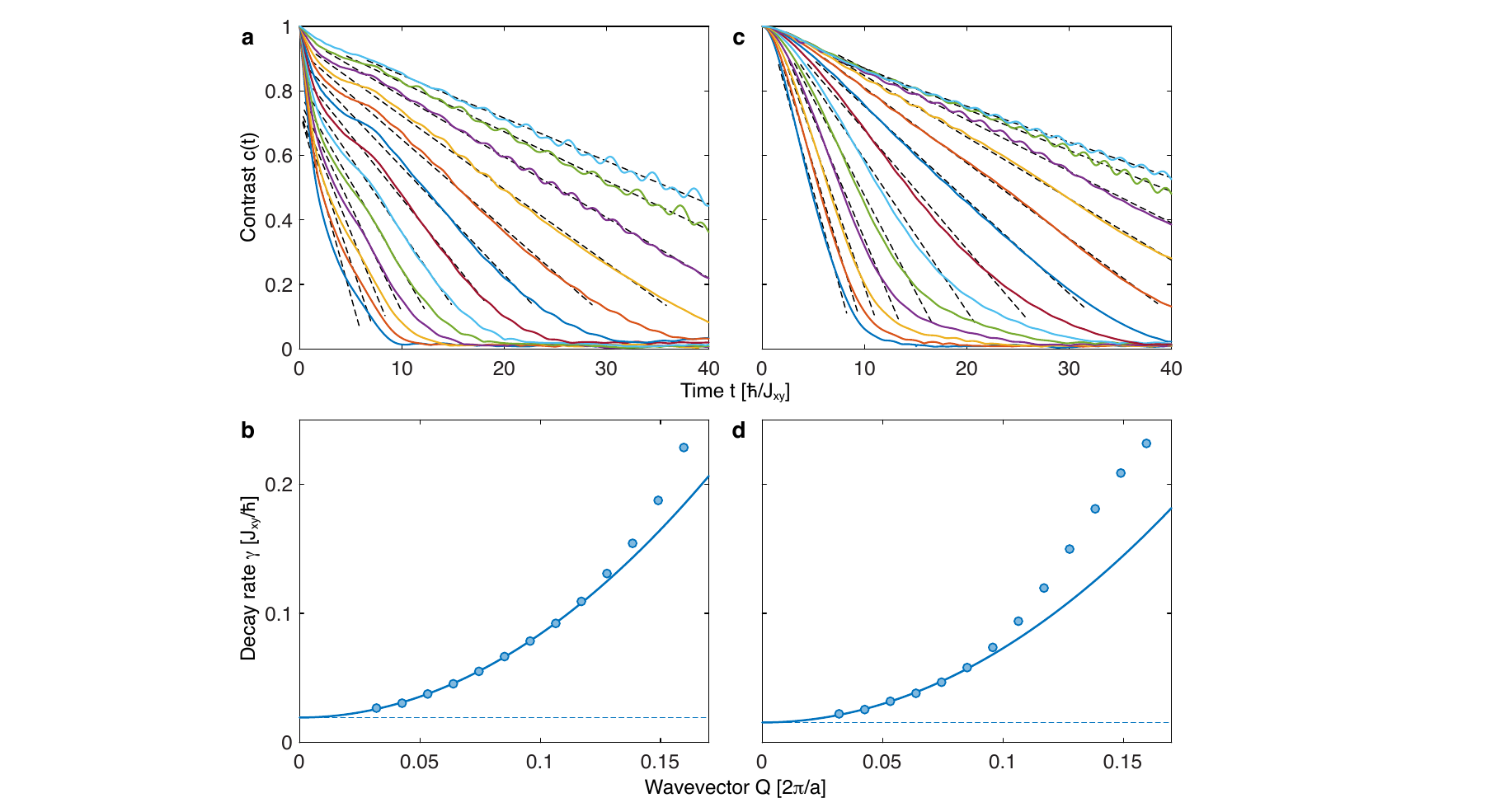} 
	\caption{\textbf{Spin relaxation for the XXZ model}. \textbf{a},\textbf{b}, Isotropic model $\Delta\,{=}\,1$ with finite hole concentration ($p\,{=}\,0.1$). \textbf{c},\textbf{d}, Slightly anisotropic $\Delta\,{=}\,0.93$ with no holes ($p\,{=}\,0$). The colored solid lines in \textbf{a} and \textbf{c} are decay curves $c(t)$ for different wavelengths $\lambda\,{=}\,31.3\,a$, $23.5\,a$, $18.8\,a$, $15.7\,a$, $13.4\,a$, $11.7\,a$, $10.4\,a$, $9.4\,a$, $8.5\,a$, $7.8\,a$, $7.2\,a$, $6.7\,a$, $6.3\,a$ (from top to bottom). The dotted lines are linear fits to determine decay rates, which are shown in \textbf{b} and \textbf{d} with a fit $\gamma = DQ^2 + \gamma_0$ (solid line). }
	\label{fig:XXZ_theory}
	\vspace{-0pt}
\end{figure*}

We consider $\Delta = 0.93$, $h_z = 0.89 J_{xy}$ (thus, simulating the experimental conditions at $11E_R$), as well as $J_{\uparrow \uparrow}/J_{\uparrow \downarrow} = 0.53$, $J_{\downarrow \downarrow}/J_{\uparrow \downarrow} = 1.41$. Figure~\ref{fig:XXZ_theory}c-d show the results for a simulation in the absence of holes. We use the same method as for the XXX model to determine decay rates from linear fits. We obtain a $Q$-independent decay rate $\gamma_0\,{=}\,0.015\,J_{xy}/\hbar$ and a \enquote{diffusion constant} $D\,{=}\,0.15\,a^2/(\hbar/J_{xy})$.
 
\subsection{Dephasing from edge effects}
\label{sec:dephasing_edge}

We also use numerical simulations   to study how the inhomogeneity of the effective magnetic field at the ends of finite chains leads to dephasing for transverse spin components.  We concentrate on the $Q\,{\to}\,0$ limit, i.e.~a state  uniformly polarized in the $x$-direction, with anisotropy $\Delta\,{=}\,1$. 

\begin{figure}[t]
	\includegraphics[width=\linewidth,keepaspectratio]{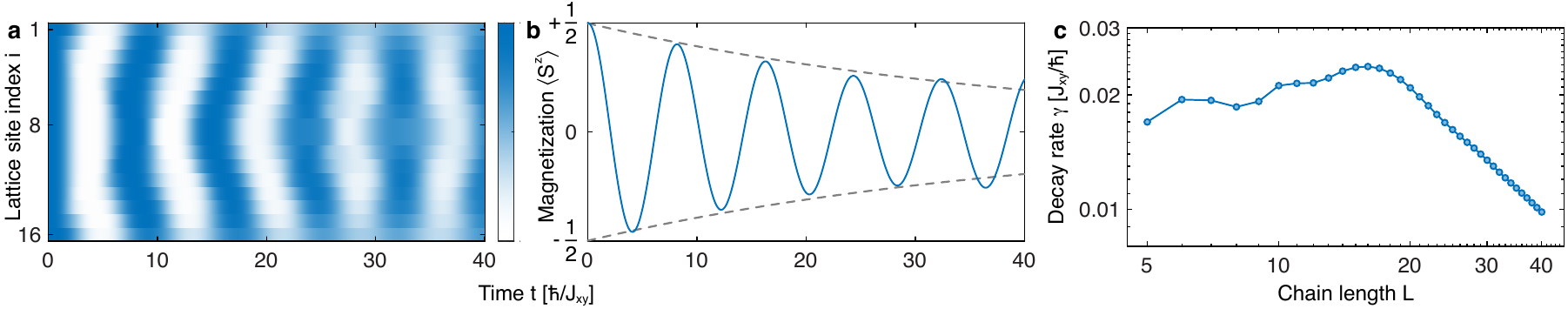} 
	\caption{\textbf{Dephasing from the edges of the chains}. \textbf{a},\textbf{b}, Dynamics of an $x$-polarized ($Q\,{=}\,0$) state for a $L\,{=}\,16$ finite chain, with anisotropy $\Delta\,{=}\,1$ and $h_z\,{=}\,0.8\,J_{xy}$ reflecting the parameters of the experiment. Panel \textbf{a} shows how the initially homogeneous phase of the locally-measured transverse spin $\langle S^x_i(t)\rangle$ gets distorted at the edges, and how this perturbation propagates through the chain, leading to a loss of overall contrast of the total transverse magnetization $\sum_i \langle S^x_i(t)\rangle$ (\textbf{b}). A fit (dashed lines) to   $\cos(\omega t  + \phi) e^{-\gamma t}/2$ determines the decay rate $\gamma$ which is shown as a function of system size $L$ in panel \textbf{c}. A log-log plot shows the scaling $\gamma\,{\propto}\,L^{-1}$ for large $L$. }
	\label{fig:theory_edge}
	\vspace{-0pt}
\end{figure}

If the effective magnetic field were globally uniform, the transverse magnetization $\sum_i \langle S^x_i(t)\rangle$ will just oscillate in time without decaying. However, the fact that the edges of the chain feel an effective magnetic field strength which is half that of the bulk, causes dephasing of spins at the edges. This disturbance in turn propagates into the bulk (see Fig.~\ref{fig:theory_edge}a), so that the transverse magnetization decays in time (see Fig.~\ref{fig:theory_edge}b). For long chains, the decay rate decreases as a function of length $L$, as the bulk dominates the edges.  This trend starts only for chains with $L\,{>}\,16$. For smaller $L$, the trend is reversed due to few-body dynamics.

We have also explored the effect of the effective magnetic field in the XX model.  Comparison of simulations of the pure spin model with and without an effective magnetic field of $h_z\,{=}\,1.43\,J_{xy}$ shows that the edge effect for chains of length $L\,{=}\,40\,a$ give rise to a $Q$-independent decay rate of $0.04\,\hbar/J_{xy}$, which would amount to shifting the decay curve shown as a black line Fig.~\ref{fig:decayCurves}b (for $h_z\,{=}\,0$) vertically. 

\section{Derivation of the bosonic $\tilde{t}$-$J$ model}
\label{Appendix:tJmodel}
We   derive here the bosonic $\tilde{t}$-$J$ model 
\begin{align}
    H_{\tilde{t}\text{-}J} & = \sum_{\langle i j \rangle} \left[ J_{xy} (S^x_i S^x_j + S^y_i S^y_j) + J_z S^z_i S^z_j -  \frac{h_z}{2} \left( S^z_i(n_{\uparrow j} +  n_{\downarrow j}) + (n_{\uparrow i } + n_{ \downarrow i}) S^z_j \right)  + c (n_{\uparrow i} + n_{\downarrow i })(n_{\uparrow j} + n_{\downarrow j})  \right]  \nonumber \\
    & - \sum_{\sigma, \langle ij \rangle} \tilde{t} a_{\sigma i}^\dagger a_{\sigma j} + \text{h.c.}   - \sum_{\sigma,\langle ijk\rangle } \left[ \frac{\tilde{t}^2}{U_{\uparrow \downarrow}} a^\dagger_{\sigma i} n_{\bar{\sigma} j} a_{\sigma k} + \frac{\tilde{t}^2}{U_{\uparrow \downarrow}} a_{\bar\sigma i}^\dagger S^\sigma_j a_{\sigma k} + \frac{2\tilde{t}^2}{U_{\sigma \sigma}} a_{\sigma i}^\dagger n_{\sigma j} a_{\sigma k} \right] + \text{h.c.}
   \label{eqn:tJ_appendix}
\end{align}
quoted in the main text. Here spin ${\sigma}\,{=}\,{\uparrow,\downarrow}$, and $a_{\sigma i}$, $a^\dagger_{\sigma i}$ are bosonic lowering and raising operators at site $i$, such that $S^\uparrow_i \equiv S^+_i := a^\dagger_{\uparrow i} a_{\downarrow i}$, $S^\downarrow_i \equiv S^-_i = (S^+_i)^\dagger$, and  $S^x_i = \frac{1}{2}(S^+_i + S^-_i)$, $S^y = \frac{1}{2}(-i S^+_i + i S^-_i)$, $S^z = \frac{1}{2}(n_{\uparrow i} - n_{\downarrow i})$ form a representation of the Pauli algebra   (to be more precise it is the direct sum of  $1/2 \oplus 0$ irreducible representations). Also $c = -\tilde{t}^2(1/U_{\uparrow \uparrow} + 1/U_{\downarrow \downarrow} + 1/U_{\uparrow \downarrow} )$.

The starting point is the  Bose-Hubbard Hamiltonian describing cold atoms moving in a deep optical lattice  (so that they are confined to the lowest Bloch band):
\begin{align}
    H = -\sum_{\langle ij \rangle, \sigma} \tilde{t} ( a^\dagger_{i\sigma} a_{j \sigma} + \text{h.c.} )  + \frac{1}{2} \sum_{i, \sigma} U_{\sigma \sigma} n_{i \sigma} (n_{i \sigma} -1) + U_{\uparrow\downarrow} \sum_i n_{i\uparrow} n_{i \downarrow}.
\end{align}
We assume two components $\sigma = \uparrow,\downarrow$ and $U_{\sigma \sigma'} \gg \tilde{t}$, such that there are at most singly occupied sites. We will derive the effective model in this limit. We do not assume $U_{\sigma \sigma'}$ are necessarily equal between themselves.
  
We employ the expansion detailed in  \cite{PhysRevX.10.021032} where it was shown how having multiple emergent $U(1)$ charges emerge in an effective Hamiltonian.  The general set-up is as such: let $\Gamma_1, \cdots, \Gamma_m$ be $m$   mutually commuting operators with integer eigenvalue spacings, and consider the Hamiltonian
\begin{align}
    H = \vec{\omega} \cdot \vec{\Gamma} + V
\end{align}
where $V$ need not commute with $\Gamma_i$. In the limit of large $|\vec{\omega}|$ we can derive an effective Hamiltonian
\begin{align}
    H_\text{eff} = \vec{\omega} \cdot \vec{\Gamma} + H'_{\vec{0}} + \frac{1}{2} \sum_{\vec{n} \neq \vec{0}} \frac{[H'_{\vec{n}}, H'_{\vec{-n}}]}{\vec{n} \cdot \vec{\omega}}  + \cdots
    \label{eqn:effH}
\end{align}
(This turns out to be the so-called van Vleck expansion). The effective Hamiltonian has emergent symmetries $[H_\text{eff}, \Gamma_i] = 0$, i.e.~the Hamiltonian is symmetric with respect to the $m$ $U(1)$ charges $\Gamma_i$.
Here $H'_{\vec{n}}$ is the $\vec{n}$-th Fourier mode of the `interaction' Hamiltonian defined on the $m$-torus $\mathbb{T}^m$:
\begin{align}
    H'(\vec\theta) & = U_0^\dagger(\vec\theta) V U_0(\vec\theta) \\
    U_0(\vec\theta) & = e^{-i \vec\theta \cdot \vec\Gamma} \\
     H'_{\vec n} &= \frac{1}{(2\pi)^m} \int_{\mathbb{T}^m} d^m\vec\theta e^{-i \vec{n} \cdot \vec{\theta}} H'(\vec\theta)
\end{align}

Applying this formalism to the Bose-Hubbard model, we note that interactions there consist of three kinds:
\begin{align}
    & U_{\downarrow \downarrow} \Gamma_1; \qquad \Gamma_1 = \sum_i \frac{1}{2} n_{i \downarrow} (n_{i\downarrow}-1)  \\
     &U_{\uparrow \uparrow}  \Gamma_2; \qquad \Gamma_2 = \sum_i \frac{1}{2} n_{i \uparrow} (n_{i\uparrow}-1)\\
     &U_{\uparrow \downarrow } \Gamma_3; \qquad \Gamma_3 = \sum_i n_{i \uparrow } n_{i \downarrow}
\end{align}
and   that $\Gamma_i$ have integer-eigenvalues and mutually commute. We therefore identify $\omega_1 = U_{\downarrow \downarrow}, \omega_2 = U_{\uparrow \uparrow}, \omega_3 = U_{\uparrow \downarrow}$. 
We define
\begin{align}
    U_0(\vec\theta) := \exp(-i \vec\theta \cdot  \vec\Gamma)
\end{align}
which gives us
\begin{align}
    H'(\vec\theta) := U_0(\vec\theta)^\dagger \sum_{\langle ij \rangle, \sigma} \tilde{t} ( a^\dagger_{i\sigma} a_{j \sigma} + \text{h.c.} ) U_0(\vec\theta).
\end{align}

We have
\begin{align}
    e^{i \theta_3 \Gamma_3} a_{i,\uparrow}^\dagger e^{-i\theta_3 \Gamma_3}  &  = a_{i,\uparrow}^\dagger + (i \theta_3) [ n_{i\uparrow} n_{i\downarrow}, a_{i,\uparrow}^\dagger ] + \frac{(i \theta_3)^2}{2!} [ n_{i\uparrow} n_{i\downarrow}, [ n_{i\uparrow} n_{i\downarrow}, a_{i,\uparrow}^\dagger ]  ] + \cdots \nonumber \\
    & = a_{i,\uparrow}^\dagger + (i \theta_3) a_{i,\uparrow}^\dagger n_{i \downarrow}  + \frac{(i \theta_3)^2}{2!} a_{i, \uparrow}^\dagger ( n_{i\downarrow}^\dagger)^2 + \cdots \nonumber \\
    & = a_{i,\uparrow}^\dagger e^{i \theta_3 n_{i \downarrow}}.
\end{align}
Therefore 
\begin{align}
& e^{i \theta_3 \Gamma_3} a_{i,\uparrow}^\dagger e^{-i\theta_3 \Gamma_3}   = a_{i,\uparrow}^\dagger e^{i \theta_3 n_{i \downarrow}}, \\
    & e^{i \theta_3 \Gamma_3} a_{i,\uparrow} e^{-i\theta_3 \Gamma_3} = a_{i,\uparrow} e^{-i \theta_3 n_{i \downarrow}}, \\
    & e^{i \theta_3 \Gamma_3} a_{i,\downarrow}^\dagger e^{-i\theta_3 \Gamma_3} = a_{i,\downarrow}^\dagger e^{i \theta_3 n_{i \uparrow}}, \\
     & e^{i \theta_3 \Gamma_3} a_{i,\downarrow} e^{-i\theta_3 \Gamma_3} = a_{i,\downarrow} e^{-i \theta_3 n_{i \uparrow}}.
\end{align}

Next 
\begin{align}
    e^{i \theta_2 \Gamma_2} a_{i,\uparrow}^\dagger e^{-i\theta_2 \Gamma_2}   & = a_{i,\uparrow}^\dagger + (i \theta_2) [ \frac{1}{2} n_{i\uparrow}(n_{i\uparrow}-1), a_{i,\uparrow}^\dagger ] + \frac{(i \theta_2)^2}{2!} [ \frac{1}{2} n_{i\uparrow}(n_{i\uparrow}-1), [ \frac{1}{2} n_{i\uparrow}(n_{i\uparrow}-1), a_{i,\uparrow}^\dagger ]  ] + \cdots \nonumber \\
\end{align}
A single commutator yields
\begin{align}
    [ \frac{1}{2} n_{i\uparrow}(n_{i\uparrow}-1), a_{i,\uparrow}^\dagger ] & = \frac{1}{2} [n_{i,\uparrow}, a_{i,\uparrow}^\dagger] (n_{i\uparrow}-1) + \frac{1}{2} n_{i,\uparrow} [(n_{i\uparrow}-1), a_{i,\uparrow}^\dagger] \nonumber \\
    & = \frac{1}{2} a_{i,\uparrow}^\dagger (n_{i\uparrow}-1) + \frac{1}{2} n_{i,\uparrow} a_{i,\uparrow}^\dagger \nonumber \\
    & =  a_{i,\uparrow}^\dagger n_{i\uparrow}
\end{align}
so the full expression becomes
\begin{align}
    e^{i \theta_2 \Gamma_2} a_{i,\uparrow}^\dagger e^{-i\theta_2 \Gamma_2}   & = a_{i,\uparrow}^\dagger + (i \theta_2) a_{i,\uparrow}^\dagger n_{i \uparrow} + \frac{(i \theta_2)^2}{2!} a_{i,\uparrow}^\dagger (n_{i \uparrow} )^2 + \cdots \nonumber \\
    & = a_{i,\uparrow}^\dagger e^{i \theta_2 n_{i\uparrow}}.
\end{align}
Similarly we have
\begin{align}
    e^{i \theta_2 \Gamma_2} a_{i,\uparrow} e^{-i\theta_2 \Gamma_2} = e^{-i \theta_2 n_{i\uparrow}} a_{i,\uparrow}.
\end{align}

Therefore, there are four terms in $H'(\vec\theta)$: 
\begin{align}
    & U_0(\vec\theta)^\dagger a_{i,\uparrow}^\dagger a_{j,\uparrow} U_0(\vec\theta) = a_{i,\uparrow}^\dagger e^{i (\theta_3 n_{i \downarrow} + \theta_2 n_{i \uparrow}) }
    e^{-i (\theta_3 n_{j,\downarrow} + \theta_2 n_{j, \uparrow} )}
    a_{j,\uparrow} \\ 
    & U_0(\vec\theta)^\dagger a_{j,\uparrow}^\dagger a_{i,\uparrow}  U_0(\vec\theta) = 
    a_{j,\uparrow}^\dagger
    e^{i (\theta_3 n_{j \downarrow} + \theta_2 n_{j \uparrow}) }
    e^{-i (\theta_3 n_{i,\downarrow} + \theta_2 n_{i, \uparrow} )}
    a_{i,\uparrow}  \\
    & U_0(\vec\theta)^\dagger a_{i,\downarrow}^\dagger a_{j,\downarrow} U_0(\vec\theta) = a_{i,\downarrow}^\dagger e^{i (\theta_3 n_{i \uparrow} + \theta_1 n_{i \downarrow}) }
    e^{-i (\theta_3 n_{j,\uparrow} + \theta_1 n_{j, \downarrow} )}
    a_{j,\downarrow} \\
    & U_0(\vec\theta)^\dagger a_{j,\downarrow}^\dagger a_{i,\downarrow}       U_0(\vec\theta) =  
    a_{j,\downarrow}^\dagger
    e^{i (\theta_3 n_{j \uparrow} + \theta_1 n_{j \downarrow}) }
    e^{-i (\theta_3 n_{i,\uparrow} + \theta_1 n_{i, \downarrow} )}
    a_{i,\downarrow} 
\end{align}

Now, the $\vec{n}$-th Fourier mode of $H'(\vec\theta)$ enforces projectors of certain occupation numbers between sites. For example for the first term and $n_1 = 0, n_2 = 0, n_3 = 1$ this is
\begin{align}
    a^\dagger_{i,\uparrow} \mathbb{P}_{n_{i,\downarrow} - n_{j,\downarrow}  = 1} \mathbb{P}_{n_{i,\uparrow} - n_{j,\uparrow}  = 0} a_{j, \uparrow}.
\end{align}

We evaluate Eq.~\eqref{eqn:effH} and the result is Eq.~\eqref{eqn:tJ_appendix} (ignoring the constant term $\vec\omega \cdot \vec\Gamma$). 


\begin{thebibliography}{10}
\bibitem{footnotePrefactor}
\bibinfo{note}{The prefactor 1.14 is somewhat arbitrarily obtained by
  integrating the probability for a localized particle to be on the original
  site, $J_0(t/ (\hbar /2\protect \mathaccentV {tilde}07E{t}))^2$ to the first
  zero of the Bessel function $J_0$}.
\end{thebibliography}
\end{document}